\documentclass[
reprint,
superscriptaddress,
amsmath,
amssymb,
aps,
pra,
floatfix,
longbibliography
]{revtex4-1}

\usepackage{braket}
\usepackage{dcolumn}
\usepackage{bm}

\usepackage[dvips]{graphicx}
\usepackage{subfigure}
\usepackage{amsmath}
\usepackage{amssymb}
\usepackage{bm}
\usepackage[colorlinks=true,citecolor=blue,urlcolor=blue,linkcolor=blue,hyperfigures=true]{hyperref}
\usepackage{multirow}
\usepackage{comment}
\usepackage{soul}
\usepackage[dvipsnames]{xcolor}

\graphicspath{{./Figures/}}

\def\Eq{Eq.~}
\def\Eqs{Eqs.~}
\def\Fig{Fig.~}

\def\Ref{Ref.~}
\def\Refs{Refs.~}

\def\dd{\mathrm{d}}

\begin{document}

\preprint{APS/123-QED}

\title{Navigation-compatible hybrid quantum accelerometer using a Kalman filter}

\author{Pierrick Cheiney}
\email[email: ]{pierrick.cheiney@ixblue.com}
\affiliation{iXblue, 34 rue de la Croix de Fer, 78105, Saint-Germain-en-Laye, France}
\affiliation{LP2N, Laboratoire de Photonique Num\'{e}rique et Nanosciences, Institut d'Optique
Graduate School, rue Fran\c{c}ois Mitterrand, 33400, Talence, France}
\author{Lauriane Fouch\'e}
\altaffiliation[Present address: ]{CEA CESTA, 15 avenue des Sabli\`eres, 33114, Le Barp, France.}
\affiliation{LP2N, Laboratoire de Photonique Num\'{e}rique et Nanosciences, Institut d'Optique
Graduate School, rue Fran\c{c}ois Mitterrand, 33400, Talence, France}
\author{Simon Templier}
\affiliation{iXblue, 34 rue de la Croix de Fer, 78105, Saint-Germain-en-Laye, France}
\affiliation{LP2N, Laboratoire de Photonique Num\'{e}rique et Nanosciences, Institut d'Optique
Graduate School, rue Fran\c{c}ois Mitterrand, 33400, Talence, France}
\author{Fabien Napolitano}
\affiliation{iXblue, 34 rue de la Croix de Fer, 78105, Saint-Germain-en-Laye, France}
\author{Baptiste Battelier}
\affiliation{LP2N, Laboratoire de Photonique Num\'{e}rique et Nanosciences, Institut d'Optique
Graduate School, rue Fran\c{c}ois Mitterrand, 33400, Talence, France}
\author{Philippe Bouyer}
\affiliation{LP2N, Laboratoire de Photonique Num\'{e}rique et Nanosciences, Institut d'Optique
Graduate School, rue Fran\c{c}ois Mitterrand, 33400, Talence, France}
\author{Brynle Barrett}
\affiliation{iXblue, 34 rue de la Croix de Fer, 78105, Saint-Germain-en-Laye, France}
\affiliation{LP2N, Laboratoire de Photonique Num\'{e}rique et Nanosciences, Institut d'Optique
Graduate School, rue Fran\c{c}ois Mitterrand, 33400, Talence, France}

\date{\today}

\begin{abstract}
Long-term inertial navigation is currently limited by the bias drifts of gyroscopes and accelerometers. Ultra-stable cold-atom interferometers offer a promising alternative for the next generation of high-end navigation systems. Here, we present an experimental setup and an algorithm hybridizing a stable matter-wave interferometer with a classical accelerometer. We use correlations between the quantum and classical devices to track the bias drift of the latter and form a hybrid sensor. We apply the Kalman filter formalism to obtain an optimal estimate of the bias and simulate experimentally a harsh environment representative of that encountered in mobile sensing applications. We show that our method is more precise and robust than traditional sine-fitting methods. The resulting sensor exhibits a $400~\mathrm{Hz}$ bandwidth and reaches a stability of $10~\mathrm{ng}$ after $11~\mathrm{h}$ of integration.
\end{abstract}

\maketitle

Inertial navigation systems determine the position of a moving vehicle by continuously measuring its acceleration and rotation rate, and subsequently integrating the equations of motion \cite{Titterton2004}. These systems are limited by slow drifts of the biases inherent to their inertial sensors, which ultimately lead to large speed and position errors after integration. Currently, the long-term bias stability of navigation-grade accelerometers is on the order of 10 $\mu$g---which, in the absence of aiding sensors such as satellite navigation systems, leads to horizontal position oscillations of 60 m at the characteristic Schuler period of $84.4$ minutes \cite{Titterton2004, Schuler1923}.

Since their first demonstration in the early 1990s, atom interferometers (AIs) have proven to be excellent absolute inertial sensors---having been exploited as ultra-high sensitivity instruments for fundamental tests of physics \cite{Bouchendira2011, Cronin2015, LZhou2015, Kovachy2015, Barrett2016b, Rosi2017}, and as state-of-the-art gravimeters with accuracies in the range of $1-10~\mathrm{ng}$ achieved both in laboratories \cite{Peters2001, LeGouet2008, Altin2013, Gillot2014, Freier2016, Hardman2016} and with compact transportable systems \cite{Bodart2010, Barrett2016c,Bidel2018,AOsense,muquans}. As a result, they have been proposed for the next generation of inertial navigation systems \cite{Jekeli2005, Geiger2011, Barrett2016a, Battelier2016}. However, cold-atom-based sensors generally possess a small bandwidth, and suffer from low repetition rates (with the exceptions of \Refs \cite{Rakholia2014,Dutta2016}) and dead times during which no inertial measurements can be made. In comparison, mechanical accelerometers exhibit broad bandwidths compatible with navigation applications \footnote{Industry standards are 200 Hz for naval application and 2 kHz for aviation}, but are afflicted by long-term bias and scale factor drifts. These two types of sensors can thus be hybridized \cite{Lautier2014} in order to benefit from the best of both worlds---in strong analogy with the strategy employed in atomic clocks \cite{Ludlow2015}.

Here, we use correlations between an AI and a classical accelerometer to track the bias of the latter, and we present an approach based on a non-linear Kalman filter (KF) \cite{Kalman1960, Kalman1961, BarShalom2002, vanTrees2013, Brown2012} to optimally track all of the interference fringe parameters---making the estimation of the accelerometer bias robust against variations of experimental parameters. We show that the hybridization procedure acts as a first-order high-pass filter on the errors of the mechanical sensor, effectively removing slow bias drifts. We simulate a mobile environment in the laboratory by adding simultaneously vibration noise, temperature variations and laser intensity fluctuations. Even under these conditions, we are able to track the accelerometer bias to less than $1~\mathrm{\mu g}$. In a typical laboratory environment, our hybrid accelerometer reaches a precision of $10~\mathrm{ng}$ after 11 hours of integration.

\begin{figure}[!t]
  \centering
  \includegraphics[width=0.48\textwidth]{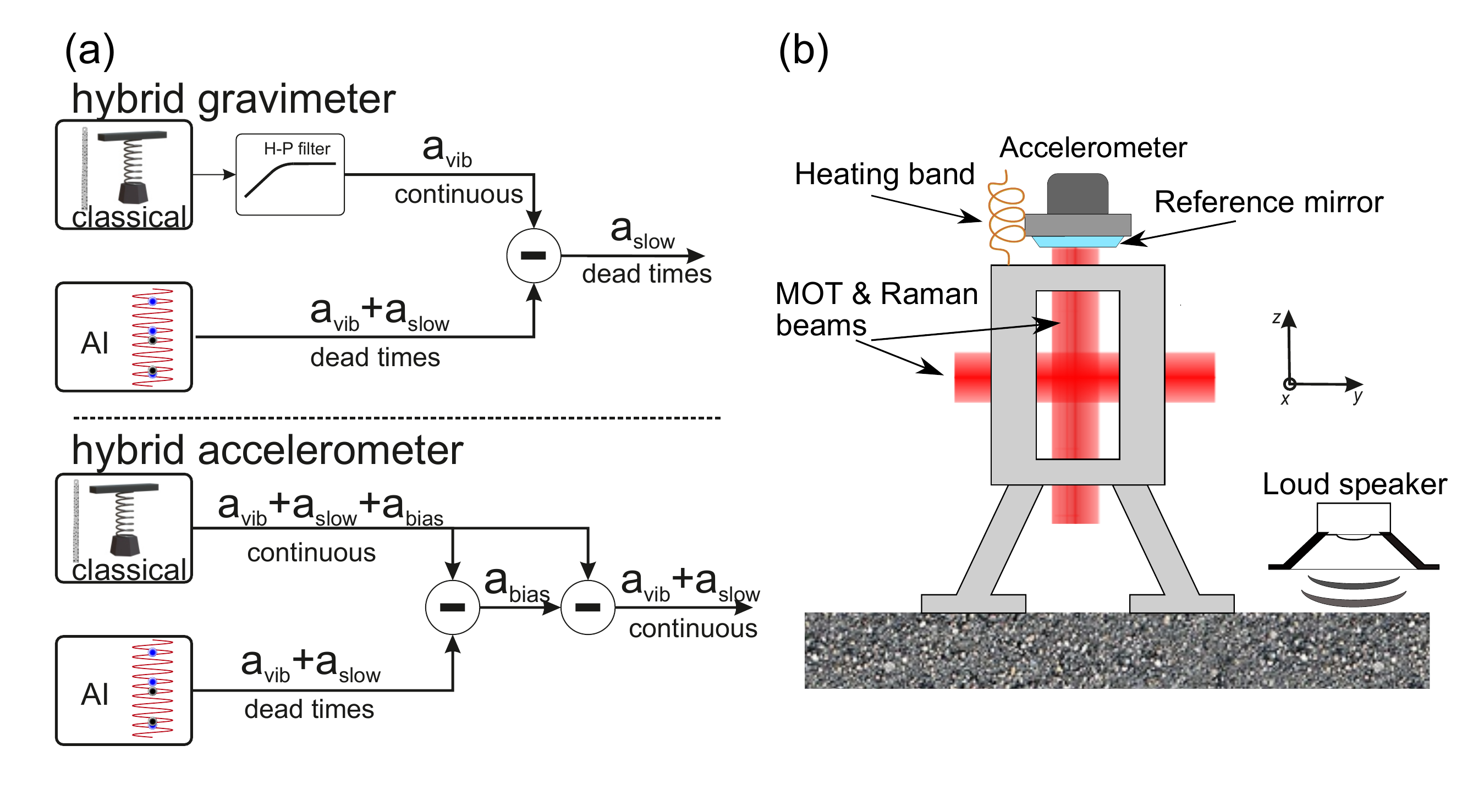}
  \caption{(a) Hybridization strategy. Correlations between classical and quantum accelerometers can be used to isolate the slowly-varying part of the acceleration (hybrid gravimeter) or to determine and subsequently reject the bias of the mechanical accelerometer (hybrid accelerometer). (b) Sketch of the experimental setup. The AI measures the free-fall acceleration of the atoms relative to the reference mirror, whose acceleration is simultaneously recorded by a mechanical accelerometer. The Raman and MOT beams share the same optical path. Heating bands are used to control the accelerometer temperature and a loud speaker is used to generate vibration noise.}
  \label{fig:Schematics}
\end{figure}

Figure \ref{fig:Schematics}(a) presents the hybridization strategy. The classical and quantum accelerometers measure acceleration simultaneously and the correlation between them is used to isolate different parts of the acceleration. By applying a high-pass filter to the classical accelerometer, the AC acceleration can be substracted from the atom interferometer output to create a hybrid gravimeter only sensitive to slow variations of the acceleration. This method has been used to digitally reject vibrations and improve the sensitivity of atom gravimeters in noisy environments \cite{Merlet2009, Geiger2011, Lautier2014, Barrett2015, Barrett2016a}. Without this filtering step, the correlations can be washed out by drifts of the classical accelerometer bias during the measurement. For navigation applications however, the DC part of the acceleration also contains relevant information. Correlations between the atom interferometer (whose bias drift is negligible) and the classical accelerometer can then be used to track the bias drifts of the latter. This can be accomplished even in a moving apparatus with non-zero mean acceleration. A continuous high-bandwidth hybrid accelerometer is then obtained by subtracting the acceleration bias from the continuous output of the classical accelerometer.

Our setup is presented in \Fig \ref{fig:Schematics}(b). It consists of a $^{87}\mathrm{Rb}$ Mach-Zender interferometer sensitive to the vertical component of acceleration. Every $1.25~\mathrm{s}$, we load $\sim 10^{9}$ atoms from background vapor into a 3D magneto-optical trap and apply standard optical molasses techniques to cool the sample to $4~\mathrm{\mu K}$. Atoms are then prepared in the lowest magnetically-insensitive state $\ket{F=1,m_F=0}$, and are subjected to a $\pi/2-\pi-\pi/2$ Raman pulse sequence---with each pulse separated by an interrogation time of $T = 20~\mathrm{ms}$. The choice of the interrogation time is the result of a tradeoff between the sensitivity of the interferometer, which increases as $T^2$, and the fall distance of the atom, which should remain smaller than the Raman beam diameter to permit operation in mobile environments with accelerations in the 0-2~g range. After the interferometer sequence, atoms in the two hyperfine ground states are detected separately by time-resolved fluorescence imaging. We reverse the direction of momentum transfer between two consecutive shots in order to reject direction-insensitive systematic errors. A $400$ Hz bandwidth low-noise mechanical accelerometer \footnote{Nanometrics Titan force-balance accelerometer. Measurements were realized using the 0.5 g clip range.}, attached to the back of the reference mirror, simultaneously records its acceleration. No anti-vibration system is implemented on our setup.

The output of the AI---given by the normalized atom number in the hyperfine state $\ket{F=2,m_F=0}$ after the final $\pi/2$-pulse---can be written as
\begin{equation}
  \label{eq:fringe}
  y = y_0 - \frac{C}{2} \cos{\left( \phi_{\mathrm{las}} + \phi_{\mathrm{acc}}\right)} + \delta u,
\end{equation}
where $y_0$ is the offset, $C$ the contrast, $\delta u$ the detection noise, $\phi_{\mathrm{las}}$ the laser phase (a control parameter), and $\phi_{\mathrm{acc}}$ the true inertial phase, which is proportional to the relative acceleration between the atoms and the reference mirror. For simplicity, we have omitted phase contributions due to systematic effects. We correlate the output of the AI with the inertial phase estimated using measurements from the mechanical accelerometer, which generally suffers from a slowly-varying bias $a_\mathrm{b}$ and high-frequency noise $\delta a$. The phase estimate can then be written as
\begin{eqnarray}
  \label{FRACphase}
  \begin{split}
  \tilde{\phi}_{\mathrm{acc}} & = k_{\rm eff}\int\! f(t)\left(a+a_\mathrm{b}+\delta a\right) \dd t\\
 & = \phi_{\mathrm{acc}} + \phi_{\mathrm{b}} + \delta \varphi,
\end{split}
\end{eqnarray}
where $f(t)$ is the AI response function to accelerations \cite{Cheinet2008, Merlet2009}. The bias phase $\phi_\mathrm{b}$ is related to the accelerometer bias via $\phi_{\rm b} = S_{\rm acc} a_{\rm b} $, where $S_{\rm acc} = k_{\rm eff} \int\! f(t) \dd t \simeq k_{\rm eff} T^2$ is the scale factor of the AI and $k_{\rm eff} \simeq 4\pi/\lambda$ is the effective wavevector of the Raman light with wavelength $\lambda$. A full fringe of our interferometer thus corresponds to an acceleration variation of $\sim100~\mathrm{\mu g}$. Finally, the phase estimate noise $\delta \varphi$ comprises errors due to the accelerometer's self noise, non-linearity, finite bandwidth, and imperfect mechanical coupling between it and the reference mirror.

In mobile applications, or in harsh environments, difficulties in determining the bias phase can stem from variations of the AI contrast and offset due to \textit{e.g.}~rotations, optical misalignments or vapor pressure variations. Furthermore, in the absence of real-time feedback, the vibration noise effectively randomizes the inertial phase---preventing the use of contrast-insensitive mid-fringe phase modulation schemes \cite{Merlet2009}.

Traditionally, the contrast, offset and bias phase are retrieved by performing a least-squares fit of the reconstructed fringe pattern to a sinusoidal function \cite{Peters2001, Merlet2009}. However, when these parameters are time-varying, it becomes necessary to form stacks of data to avoid washing out the fringe pattern. The choice of the number of points per stack is then associated with a trade-off between precision and bandwidth. This is characteristic of a waveform estimation problem, \textit{i.e.}~the search for the best estimator of the state of a \textit{time-varying} system.

The KF formalism provides a more elegant method that avoids this trade-off and, under reasonable assumptions, provides an optimal estimate of the fringe pattern parameters along with their full statistical properties. The KF has become a very popular estimator thanks to its simplicity and versatility, and is ubiquitous in optimal control theory \cite{Brown2012}. It is used extensively to combine different types of sensors in inertial navigation \cite{Titterton2004, Groves2013}, and has also been applied for example in optical interferometry \cite{Yonezawa2012} and more recently to track the state of an atomic magnetometer \cite{Jimenez-Martinez2018}. For linear systems driven by white Gaussian processes and observed with unbiased white Gaussian noise, the KF is an optimal estimator in the sense that it minimizes the mean-squared error of the estimation (see Appendix \ref{sec:KF demonstration}). The KF uses all previous data in an iterative way that requires very little memory and computational power---making it particularly attractive for real-time feedback and onboard applications. Even for non-linear systems, as in the present case, the KF can be linearized and provides a near-optimal solution.

The iterative KF algorithm can be split into two steps: a \textit{propagation} step, where the estimate of the tracked waveform and its covariance are updated between two measurements according to a model of the system dynamics, and a \textit{measurement} step where the latest data point is used to correct the previous estimate. Although only the previous estimate is used at each step, \emph{all} previous measurements contribute to the construction of each new estimate---whereas with sine-fitting or non-linear locking techniques \cite{Merlet2009}, this information is effectively discarded. Specifically, we model the AI fringe pattern with a four-parameter state vector
\begin{equation}
\bm{x} =
  \begin{bmatrix}
    \phi_\mathrm{b} \\
    \phi'_\mathrm{b}\\
    y_0 \\
    C
  \end{bmatrix}\!,
\end{equation}
where $\phi'_\mathrm{b}$ is the time-derivative of the bias phase $\phi_\mathrm{b}$. We model the statistical evolution of $\phi'_\mathrm{b}$, $y_0$ and $C$ with independent Wiener processes. The time evolution of the state vector is then governed by the discrete-time stochastic equation
\begin{equation}
  \label{Eq:stochDyn}
  \bm{x}(t+\delta t) = \bm{F} \cdot \bm{x}(t) + \bm{w},
\end{equation}
where $\delta t$ is the time between two consecutive measurements and is not necessarily constant, $\bm{F}$ is the evolution matrix and $\bm{w}$ is a vector of independent, normally-distributed random variables with zero mean and standard deviations $\sigma_j \delta t$ for each element $j$ of the state vector. Since these stochastic driving variables are independent, the associated covariance matrix $\bm{Q}$ contains only diagonal elements
\begin{equation}
\bm{F}=
  \begin{bmatrix}
    1 & \delta t & 0 & 0\\
    0 & 1 & 0 & 0\\
    0 & 0 & 1 & 0  \\
    0 & 0 & 0 & 1
  \end{bmatrix}\!,
\;\;\;
\bm{Q} = \delta t^2
  \begin{bmatrix}
    0 & 0 & 0 & 0\\
    0 & \sigma_{\phi'}^2 & 0 & 0\\
    0 & 0 & \sigma_{y_0}^2 & 0  \\
    0 & 0 & 0 & \sigma_{C}^2
  \end{bmatrix}\!.
\end{equation}
We emphasize that the phase is driven indirectly through its time-derivative (the top-left element of the matrix $\bm{Q}$ is zero). This permits us to optimally track a linearly varying bias phase without time-lag error, similar to the integral component of a feedback loop.

\begin{figure}[!t]
  \centering
  \includegraphics[width=0.44\textwidth]{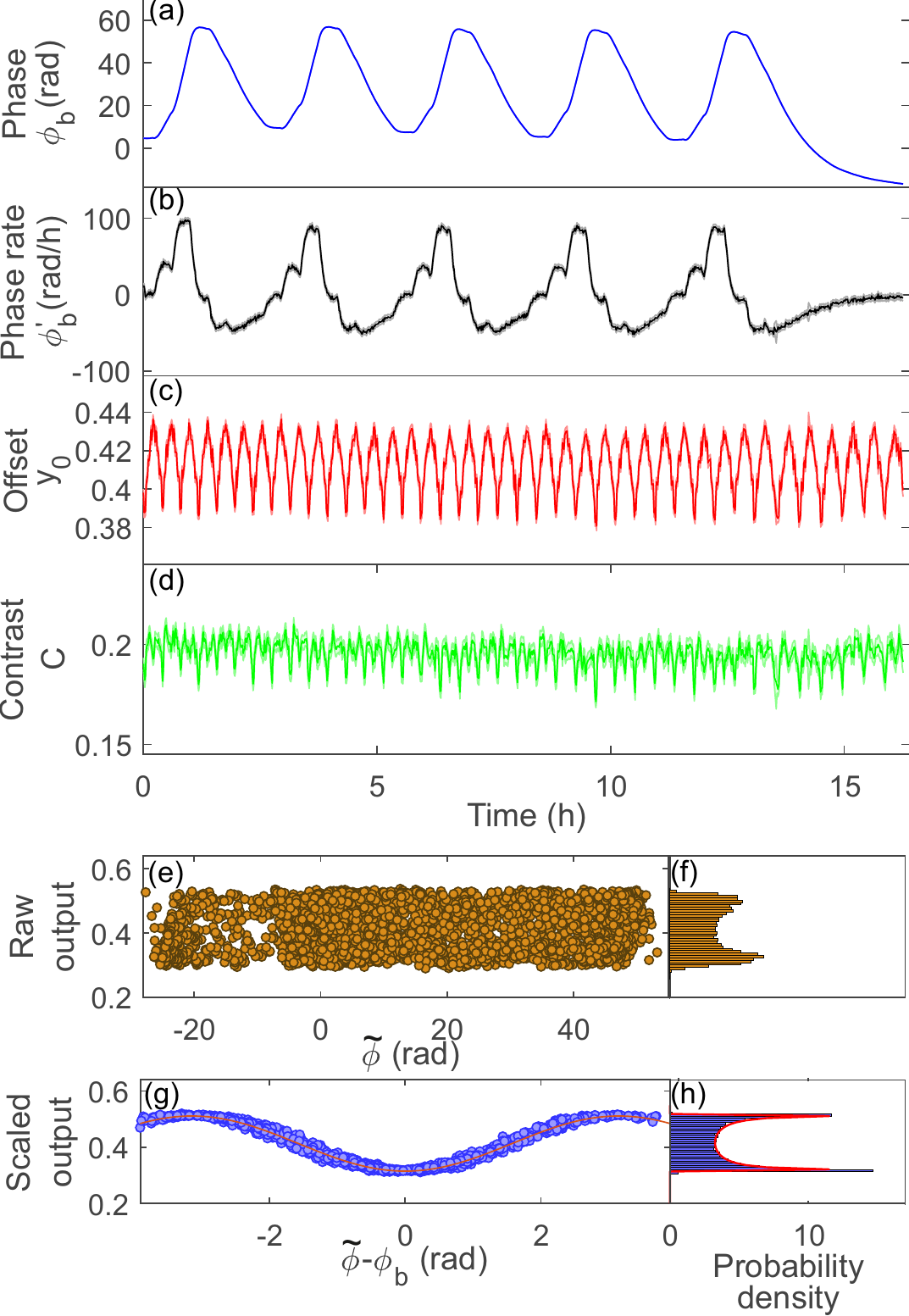}
  \caption{Tracking of the bias phase (a), bias phase rate (b), offset (c) and contrast (d) using the non-linear KF. The shaded areas correspond to the standard deviations estimated by the KF. Because of the temperature variation, the bias phase oscillates by $\sim 50$ rad during the measurement. The Raman beam intensity is modulated independently and leads to a $10\%$ variation in the offset and contrast of the interference pattern. (e) Raw AI output as a function of the estimated phase $\tilde{\phi}$ and (f) corresponding output probability distribution. (g) Scaled AI output as a function of the corrected phase estimate $\tilde{\phi}-\phi_b$ and (h) corresponding probability distribution. While this distribution is washed out in the raw data, the scaled AI output matches closely the expected arcsine distribution (solid red curve).}
  \label{fig:StateVector}
\end{figure}

In the propagation step, the pre-measurement estimate is deduced from the results of the previous post-measurement estimate
\begin{subequations}
\begin{align}
  \bm{x}_{i+1}^{-} & = \bm{F} \cdot \bm{x}_{i}^{+}, \\
  \bm{P}_{i+1}^{-} & = \bm{F} \bm{P}_{i}^{+} \bm{F}^{\rm T} + \bm{Q},
\end{align}
\end{subequations}
where the $-$ ($+$) superscripts indicate the pre- (post-) measurement estimate, and the subscript $i$ denotes the $i^{\rm th}$ measurement. The covariance matrix $\bm{P}$ characterizes the estimation uncertainty. Since the measurement process described by \Eq \eqref{eq:fringe} is a non-linear function of the state vector, we use the extended non-linear KF algorithm \cite{vanTrees2013, Brown2012}.
\begin{figure*}[!t]
  \centering
  \includegraphics[width=0.95\textwidth]{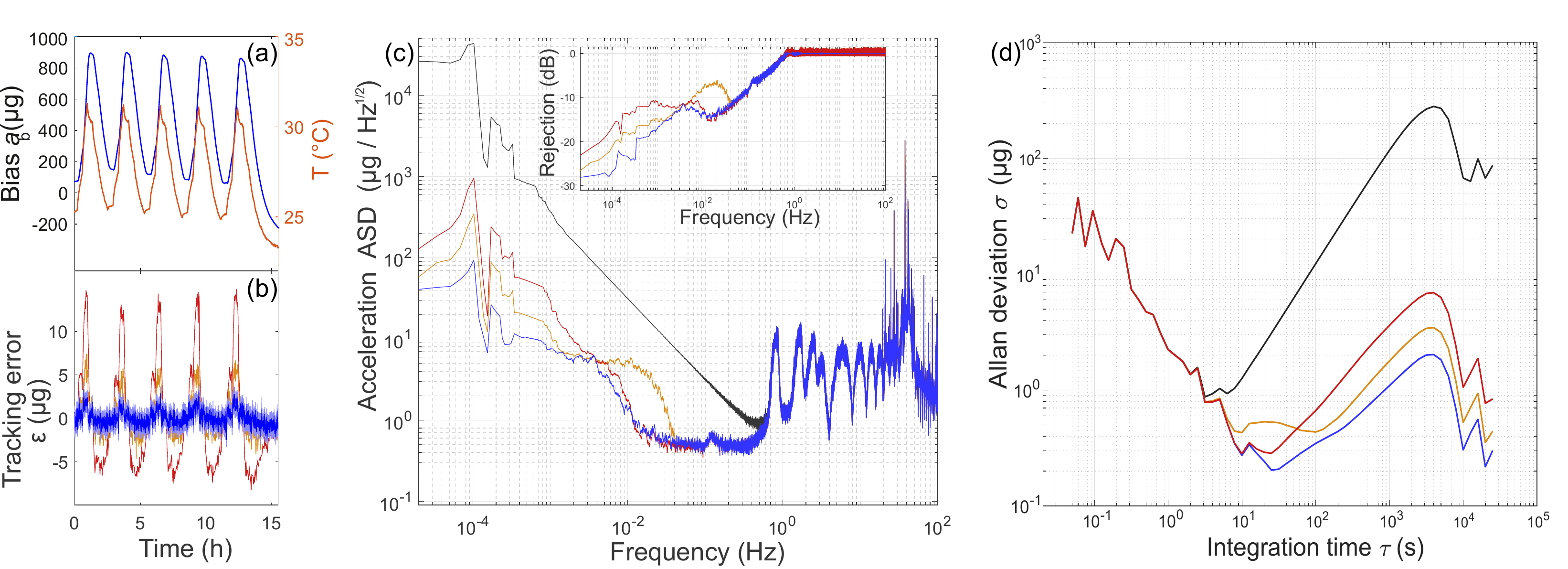}
  \caption{(a) Accelerometer bias determined by the KF algorithm (blue) and temperature (red) as a function of time. The temperature modulation produces large bias variations of $\sim 1$ mg. The standard deviation of the estimate is smaller than the line thickness. (b) Bias tracking error using the KF (blue) and by sine-fitting with 8 (brown) and 25 (red) points. The RMS value of the true bias tracking error is $\sqrt{\langle \epsilon^2_{\mathrm{KF}}\rangle} = 0.89~\mathrm{\mu g}$ for the KF---in good agreement with the estimated standard deviation, displayed as the blue shaded area. For the sine-fitting method, the true RMS errors are $\sqrt{\langle \epsilon^2_{\mathrm{SF8}}\rangle} = 2.3~\mathrm{\mu g}$ and $\sqrt{\langle \epsilon^2_{\mathrm{SF25}}\rangle} = 5.9~\mathrm{\mu g}$ for 8- and 25-point stacks. (c-d) Amplitude spectral density and Allan deviation of the standalone (black) and hybrid accelerometers using the KF (blue), and sine-fitting with 8-point (brown) and 25-point (red) stacks. The ASD shows that at low frequencies, the error rejection (inset) corresponds to a first-order high-pass filter. The error rejection is also visible in the Allan deviation, where the long term drift of the hybrid accelerometer is reduced by more than two orders of magnitude compared to the standalone one.}
  \label{fig:HybridSensor}
\end{figure*}
The trajectory is then refined after each measurement according to
\begin{subequations}
\begin{align}
  \bm{x}_i^+ & = \bm{x}_i^- + \bm{K}_i \bm{r}_i \\
  \bm{P}_i^+ & = \left(\bm{I} - \bm{K}_i \bm{H}_i\right) \bm{P}_i^-,
\end{align}
\end{subequations}
where $\bm{r}_i = \bm{y}_i - y(\bm{x}_i^{-})$ is the \textit{innovation} (\textit{i.e.}~the difference between the actual measurement and the expected output), $\bm{I}$ is the identity matrix, and the measurement matrix $\bm{H}_i = \bm{\nabla_x} y |_{\bm{x}_i}$ is the Jacobian of the AI output
\begin{equation}
  \bm{H} = \begin{bmatrix}
      \frac{C}{2} \sin(\tilde{\phi}_\mathrm{acc}-\phi_\mathrm{b}) & 0 & 1 & -\frac{1}{2}\cos(\tilde{\phi}_\mathrm{acc}-\phi_\mathrm{b})
    \end{bmatrix}.
\end{equation}
This matrix quantifies the sensitivity of the measurement to each parameter, and is calculated at each step around the estimated trajectory. Finally, the KF is optimal for the Kalman gain
\begin{equation}
  \bm{K}_i = \bm{P}_i \bm{H}_i^{\rm T} \left(\bm{H}_i \bm{P}_i \bm{H}_i^{\rm T} + \bm{R}_i \right)^{-1},
\end{equation}
where $\bm{R}_i$ is the variance of the measurement noise \footnote{The measurement noise comprises the detection noise, the phase estimation noise and the true phase noise}. We point out that the optimal Kalman gain $\bm{K}_i$ is the result of a compromise between the uncertainty of the previous estimate and the measurement noise. See the Appendices for further information regarding the KF algorithm.

We apply the KF to a 16-hour dataset where the vertical acceleration is measured by the AI and where, to simulate a mobile environment, we add the following elements [see \Fig \ref{fig:Schematics}(b)]. (\textit{i}) A loud speaker fixed to the optical table generates a $5~\mathrm{mg}$-amplitude vibration noise at $38~\mathrm{Hz}$ that randomly scans the AI phase across several fringes. (\textit{ii}) Heating bands surrounding the accelerometer are used to modulate its temperature by $\sim 5\mathrm{^{\circ}C}$ in order to induce a large bias drift ($\sim1$ mg). (\textit{iii}) The Raman beam intensity is modulated by $\sim 10\%$ using an acousto-optic modulator in the laser setup to simulate laser power fluctuations.

Figures \ref{fig:StateVector}(a-d) present the four components of the state vector $\bm{x}$ tracked by the KF. It is clear from \Fig \ref{fig:StateVector}(a) that the accelerometer bias variation corresponds to about 8 AI fringes. The step-like behavior of the heating process is clearly visible in the tracked phase rate shown in \Fig \ref{fig:StateVector}(b). The contrast and offset of the AI are also modulated by $10\%$ due to the applied Raman beam intensity modulation. The covariance matrix $\bm{P}_i$---computed by the KF algorithm at each step---provides the uncertainty of each waveform parameter. After an initial transient time of 20 s, the individual standard deviations stabilize to $\delta \phi_\mathrm{b} = 56~\mathrm{mrad}$, $\delta y_0 = 3\times 10^{-3}$ and $\delta C = 4.5\times10^{-3}$. The stabilization to a finite precision results from the competition between the amount of information provided by each measurement and the drift of the state vector. This behavior is characteristic of a waveform estimation problem \footnote{For time-invariant driving and measurement, the steady-state covariance can be determined by solving the discrete-time algebraic Riccati equations. This is not possible here because the measurement and noise matrices $\bm{H}$ and $\bm{R}$ vary randomly.}.

Figure \ref{fig:StateVector}(e) shows the AI output as a function of the estimated phase without correction. The fringe pattern is completely washed out by the bias phase variations. Similarly, the output probability distribution presented in \Fig \ref{fig:StateVector}(f) is partially smeared out by the contrast and offset variations. In comparison, well-defined scaled fringes are presented in \Fig \ref{fig:StateVector}(g), where the bias phase $\phi_\mathrm{b}$ has been subtracted from the inertial phase estimate, and the output has been scaled in a similar manner to account for the offset and contrast variations. Additionally, the scaled output probability density closely matches the expected arcsine distribution, as shown in \Fig \ref{fig:StateVector}(h). Using the scaled fringe pattern, the standard deviation of the detection and phase noise can be determined independently (see Appendix \ref{sec:Kalman optimization}). We find $\sigma_u = 2.5 \times 10^{-3}$ and $\sigma_\varphi = 0.13~\mathrm{rad}$, which indicates that the phase noise dominates and corresponds to an average sensitivity of $3.2~\mathrm{\mu g}$ per shot. This phase noise originates from the RF chain that generates the Raman frequencies, and is ultimately imprinted on the Raman lasers via electro-optic modulation.

To evaluate the precision of the bias tracking, we compare the acceleration bias estimate directly to the low-pass-filtered accelerometer output \footnote{These data are acquired by a 16-bit acquisition system at a sampling rate of 50 kHz and averaged by packets of $5~\mathrm{ms}$ for the full 16 h duration of the dataset.}. Indeed, in a static configuration, the real DC acceleration reduces to the gravitational field, which is constant to less than $10~\mathrm{ng}$ after removal of the tidal gravity anomaly. We emphasize that although we use this method to assess the quality of the tracking, the tracking itself can be performed in motion. Figure \ref{fig:HybridSensor}(a) displays the KF bias estimate along with the sensor temperature. The temperature modulation produces a large bias modulation of $\sim 1$ mg---in good agreement with the inherent temperature sensitivity of the mechanical accelerometer ($320~\mathrm{\mu g/^{\circ}C}$). The bias modulation is delayed by approximately 20 minutes compared to the temperature due to the thermal inertia of the accelerometer.

Figure \ref{fig:HybridSensor}(b) shows the acceleration bias tracking error $\epsilon$ using the KF and the sine-fitting technique with stacks of 8 and 25 points \footnote{The initial guess of each fit corresponds to the result of the previous one and the bias tracking is interpolated linearly between two consecutive stacks.}. The RMS value of the tracking error using the KF is $\sqrt{\langle \epsilon^2_{\mathrm{KF}}\rangle} = 0.89~\mathrm{\mu g}$, which agrees with the average KF standard deviation estimate $\sigma = 0.8~\mathrm{\mu g}$. These results are significantly better than the error produced by sine-fitting with stacks of 8 ($\sqrt{\langle \epsilon^2_{\mathrm{SF8}}\rangle} = 2.3~\mathrm{\mu g}$) and 25 points ($\sqrt{\langle \epsilon^2_{\mathrm{SF25}}\rangle} = 5.9~\mathrm{\mu g}$). Although strongly reduced by almost 30 dB in the KF case, the large bias modulation is still clearly visible.

We now obtain a continuous high-bandwidth hybrid sensor by subtracting the bias from the classical accelerometer output. More insight on the hybridization and the advantages of the KF can then be gained by inspecting the amplitude spectral density (ASD) of the sensors. Figure \ref{fig:HybridSensor}(c) shows the ASD of the standalone mechanical accelerometer and the hybrid accelerometer using the KF and the sine-fitting techniques. For frequencies larger than the AI cycling rate ($\sim 0.8$ Hz), the hybridization has no effect and the ASD corresponds to the vibration excitation of the reference mirror. At frequencies $f < 0.3$ Hz, the ASD of the standalone accelerometer rises reflecting the bias instability, with the main bias modulation component visible around $10^{-4}$ Hz. In comparison, for both tracking algorithms, the ASD of the hybrid sensor is reduced by several orders of magnitude at low frequencies. However, large differences can be observed between the performance of the different hybridization methods.

The inset of \Fig \ref{fig:HybridSensor}(c) shows the error rejection obtained by dividing the ASD of the hybrid accelerometer by that of the standalone one. Fitting sinusoids with stacks of 25 points performs better than with 8-point stacks by 5 dB in the 10 -- 50 mHz frequency range, but is worse by 3 dB for $f < 10~\mathrm{mHz}$. Indeed, a large number of points reduces the uncertainty of each fit---reducing the high-frequency noise, but at the expense of decreased tracking bandwidth. The KF avoids this trade-off and outperforms the sine-fitting method over the whole frequency range. In all cases, the rejection scales as $1/f$ at low frequencies---indicating that these hybridization methods can be viewed as a first-order high-pass filter of the accelerometer error.

A complementary point of view is given by the Allan deviation of the standalone and hybrid accelerometers, as shown in \Fig \ref{fig:HybridSensor}(d). The temperature modulation of the standalone unit gives rise to a large instability at integration times $\tau$ of the order of $1~\mathrm{h}$. Our hybridization strategies allow us to reduce this instability by more than two orders of magnitude. In the case of the KF hybridization, the Allan deviation does not rise above $2~\mathrm{\mu g}$, and reaches the $100~\mathrm{ng}$ level after 3 hours of integration.

\begin{figure}[!b]
  \centering
  \includegraphics[width=0.48\textwidth]{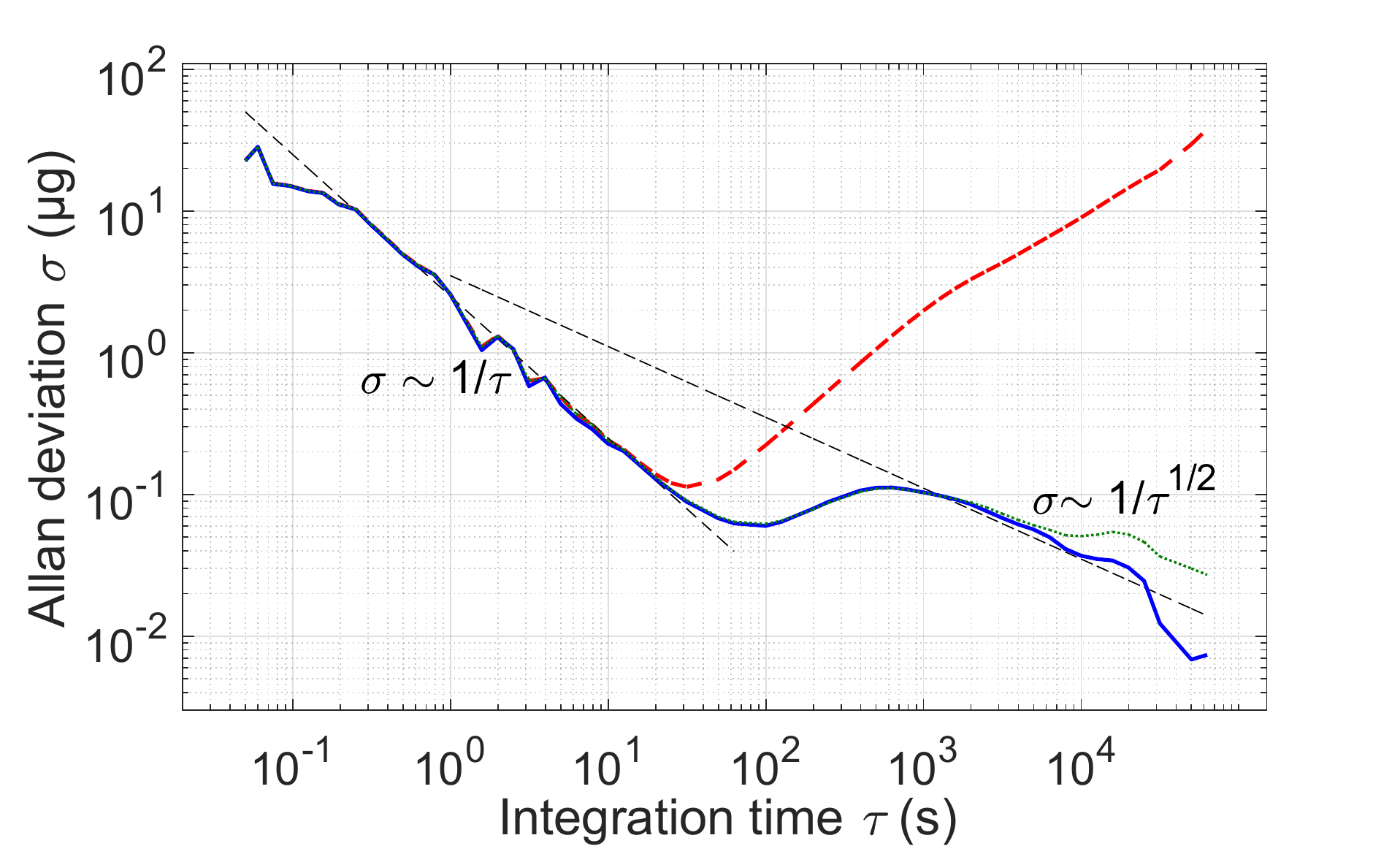}
  \caption{Allan deviation of acceleration signals from the standalone accelerometer (dashed red line) and the hybrid accelerometer using the KF with (solid blue line) and without (dotted green line) a tidal correction.}
  \label{fig:AllanDeviation}
\end{figure}

To evaluate the ultimate performance of the hybrid sensor, we record data continuously for 36 hours using a $T = 20$ ms interferometer in a typical laboratory environment, with a rms temperature stability of $\sim 0.5^\circ$ C and ambient laser power fluctuations of $\sim 1\%$. Figure \ref{fig:AllanDeviation} shows the Allan deviation of the standalone and hybrid accelerometers with and without subtraction of the tidal gravity anomaly. At small integration times, the Allan deviation of both sensors decreases as $1/\tau$, which is characteristic of averaging the sum of incommensurable periodic noises due to ambient vibrations in the laboratory. After only $30~\mathrm{s}$, the Allan deviation of the standalone accelerometer increases due to its bias instability. The Allan deviation of the hybrid sensor, however, stays below $1~\mathrm{\mu g}$ and decreases at large times as $\sim \sigma_{\mathrm{AI}}/\sqrt{\tau}$, where $\sigma_{\mathrm{AI}} = 3.2~\mathrm{\mu g/\sqrt{Hz}}$ corresponds to the AI sensitivity. For integration times larger than $5000~\mathrm{s}$, the tidal anomaly limits the Allan deviation. Nevertheless, it can be removed efficiently using an appropriate theoretical model \cite{Camp2005}. The Allan deviation then reaches a stability of $10~\mathrm{ng}$ after $4 \times 10^4~\mathrm{s}$ of integration. Up to this point we observe no signs of long-term instability in the hybrid sensor.

In conclusion, we have used a method based on the KF formalism to hybridize quantum and classical accelerometers in a simulated environment resembling that encountered in navigation applications. The hybrid sensor combines the large bandwidth and continuous measurement of a classical accelerometer with the long-term stability of a cold-atom interferometer. In addition to being more efficient computationally than least-squares sine-fitting routines, we have shown that the KF allows for a significantly more precise and robust determination of the accelerometer bias. The short-term sensitivity of the hybrid sensor is determined by the classical accelerometer noise, while the long-term stability is given by the AI. For a total interrogation time of only $2T = 40$ ms, we demonstrate a precision of $10~\mathrm{ng}$ after 11~h of integration. Such a small bias would lead to Schuler position oscillations only $60~\mathrm{mm}$ in amplitude.

For future studies, the modest interrogation times of our AI will permit operation along multiple axes \cite{Canuel2006, Wu2017} and in mobile environments with accelerations in the $0-2~\mathrm{g}$ range. The KF method presented here can also be extended to other AI configurations such as gyroscopes \cite{Gauguet2009} or gradiometers \cite{Stockton2007}, or for the differential phase extraction in dual-species tests of the equivalence principle \cite{Bonnin2013, Barrett2015, LZhou2015}. In this work we assumed that the phase and detection noise were constant in time, but extensions, such as the adaptive Kalman filter, could further improve the robustness of the bias estimate.

This work is supported by the French national agencies ANR (l'Agence Nationale pour la Recherche), DGA (D\'{e}l\'egation G\'{e}n\'{e}rale de l'Armement) under the ANR-17-ASTR-0025-01 grant, IFRAF (Institut Francilien de Recherche sur les Atomes Froids), and action sp\'{e}cifique GRAM (Gravitation, Relativit\'{e}, Astronomie et M\'{e}trologie). We would like to thank G. Condon, L. Chichet and M. Rabault for discussions during the early stages of this project, R. Jim\'enez-Mart\'{\i}nez and J-P. Michel for insightful discussions and careful reading of the manuscript, and O. Jolly for technical assistance. P. Bouyer thanks Conseil R\'{e}gional d'Aquitaine for the Excellence Chair.

\clearpage

\appendix
\section{Details on the Kalman filter algorithm}
\label{sec:KF demonstration}

In this Appendix, we present further details on the Kalman filter algorithm and demonstrate its optimality in the discrete-time linear Gaussian case. We follow closely \Ref \cite{Brown2012}, and we refer the reader to classic textbooks such as \Refs \cite{BarShalom2002, vanTrees2013} for additional information.

The filtering problem involves providing the best estimate $\hat{\bm{x}}$ of the true state vector $\bm{x}$, which is a random variable driven stochastically and measured through noisy measurements. As criteria for determining the efficiency of the filter, we use the mean-squared error of the estimation, which can be written as a matrix of covariances
\begin{equation}
  \label{eq:def_P}
  \bm{P} = \mathrm{E}[(\bm{x}-\hat{\bm{x}})(\bm{x}-\hat{\bm{x}})^{\rm T}],
\end{equation}
where $\mathrm{E}[\cdots]$ denotes the statistical expectation value. In the discrete-time case, measurements are performed at times $t_i$, labelled by the integer index $i$. Between measurements, the state vector evolves according to a linear stochastic process. The evolution of the state vector can thus be written as
\begin{equation}
  \label{eq:prop}
  \bm{x}_{i+1} = \bm{F}_i \bm{x}_i + \bm{w}_i,
\end{equation}
where $\bm{F}_i$ is the known deterministic evolution matrix and $\bm{w}_i$ is a vector of random variables. Furthermore, we assume a linear observation process such that at each time $t_i$, the measurement output $\bm{z}_i$ can be written as
\begin{equation}
  \label{eq:measurement}
  \bm{z}_i = \bm{H}_i \bm{x}_i+ \bm{v}_i,
\end{equation}
where $\bm{H}_i$ is the measurement matrix and $\bm{v}_i$ is a vector of measurement noises. Finally, we assume that both the stochastic driving and measurement noises are uncorrelated, zero-mean, white Gaussian processes such that
\begin{subequations}
\label{eq:noises}
\begin{align}
  \mathrm{E}[\bm{v}_i \bm{v}_j^{\rm T}]
  & = \delta_{ij} \bm{R}_i, \\
  \mathrm{E}[\bm{w}_i \bm{w}_j^{\rm T}]
  & = \delta_{ij} \bm{Q}_i,
\end{align}
\end{subequations}
where $\delta_{ij}$ is the Kronecker delta. Let us now assume that we have an initial estimate of the state vector and covariance immediately after an observation $\hat{\bm{x}}_i^+$, $\bm{P}_i^+$. Since the stochastic driving $\bm{w}_i$ has zero mean, the evolution of the state vector estimation between two measurements is simply
\begin{equation}
  \label{eq:prop_step}
  \hat{\bm{x}}_{i+1}^{-} = \bm{F}_i \hat{\bm{x}}_i^{+},
\end{equation}
and the estimation error is then
\begin{equation}
  \label{eq:prop_step_error}
  \bm{e}_{i+1}^- = \bm{F}_i \bm{e}_i + \bm{w}_i.
\end{equation}
Since the stochastic driving and previous estimation errors are not correlated, the error covariance evolves according to
\begin{equation}
  \label{eq:prop_step_cov}
  \bm{P}_{i+1}^-
  = \mathrm{E}[\bm{e}_{i+1}^- \bm{e}_{i+1}^{-T}]
  = \bm{F}_i \bm{P}_i^+ \bm{F}_i^{\rm T}+ \bm{Q}_i.
\end{equation}
Equations \eqref{eq:prop_step} and \eqref{eq:prop_step_cov} constitute the propagation step of the KF. In the measurement step, the new measurement is blended into the state vector linearly
\begin{equation}
  \label{eq:blend}
  \hat{\bm{x}}_i^+ = \hat{\bm{x}}_i^{-} + \bm{K}_i \left(\bm{z}_i - \bm{H}_i \hat{\bm{x}}_i^- \right),
\end{equation}
where the gain $\bm{K}_i$ is not yet determined. The KF is optimal for a particular gain $\bm{K}_i$ that minimizes the post-measurement error (\textit{i.e.}~the covariance):
\begin{equation}
  \label{eq:Pkp}
  \bm{P}_i^+ = \mathrm{E} \left[\big( \bm{x}_i - \hat{\bm{x}}_i^+ \big)\big( \bm{x}_i - \hat{\bm{x}}_i^+ \big)^{\rm T} \right].
\end{equation}
Substituting \Eqs \eqref{eq:blend} and \eqref{eq:measurement} into \eqref{eq:Pkp}, we find
\begin{align}
  \label{eq:Pkp_exp}
  \!\!\! \bm{P}_i^+
  & = \mathrm{E} \Big\{\big[(\bm{x}_i - \hat{\bm{x}}_i^-) - \bm{K}_i \big(\bm{H}_i (\bm{x}_i - \hat{\bm{x}}_i^-) + \bm{v}_i\big) \big] \nonumber \\
  & \;\;\;\; \big[ (\bm{x}_i-\bm{\hat{x}}_i^-) - \bm{K}_i \big(\bm{H}_i (\bm{x}_i - \bm{\hat{x}}_i^-) + \bm{v}_i\big) \big]^{\rm T} \Big\}, \\
  & = \big( \bm{I} - \bm{K}_i \bm{H}_i \big) \bm{P}_i^ - \big(\bm{I} - \bm{K}_i \bm{H}_i\big)^{\rm T} + \bm{K}_i \bm{R}_i \bm{K}_i^{\rm T}. \nonumber
\end{align}
The diagonal terms of the covariance $\bm{P}_i^+$ correspond to the individual errors of the state vector parameters. The KF is thus optimal for the gain $\bm{K}_i$ that minimizes these terms. Since $\bm{K}_i$ possesses enough degrees of freedom, this optimization is equivalent to simply minimizing the trace of the post-measurement covariance. By setting the derivative of Tr$[\bm{P}_i^+]$ with respect to $\bm{K}_i$ equal to zero, we find the optimal gain
\begin{equation}
  \label{eq:Kalman_gain}
  \bm{K}_i = \bm{P}_i^- \bm{H}_i^{\rm T} \big( \bm{H}_i \bm{P}_i^- \bm{H}_i^{\rm T} +\bm{R}_i \big)^{-1},
\end{equation}
which is called the Kalman gain.

\section{Kalman filter optimization}
\label{sec:Kalman optimization}

The performance of the KF relies on the accurate knowledge of the statistical properties of the stochastic driving parameters (described by the matrix $\bm{Q}$) and of the measurement noise (described by the matrix $\bm{R}$). The KF performance can be evaluated by inspecting the distribution of the \textit{innovation} $r$. For a well-tuned KF, the width of the innovation distribution is limited by the measurement noise. Therefore, it is possible to optimize the KF by minimizing the variance of the innovation, $\sigma_r^2$. However, because of compensation effects between the parameters of the noise and the stochastic dynamics, the direct minimization of $\sigma_r^2$ over all parameters often poorly estimates each parameter individually. To solve this issue, we find the optimal KF parameters in an iterative way. First, we apply the KF on the dataset using an arbitrary set of parameters. This provides an initial sub-optimal waveform estimate that is used to estimate the measurement noise. We then minimize $\sigma_r$ over the stochastic driving variables using only the estimated noise parameters to obtain a more precise estimate of the waveform. This process is then iterated a few times until a stable state is reached.

\subsection{Measurement noise estimation}
In our case, the measurement noise can be decomposed into a phase noise, which comprises the phase estimation error from the accelerometer signal and the interferometer phase noise, and a detection noise associated with the fluorescence imaging system. To highlight these noise sources, we rewrite the interferometer signal in \Eq \eqref{eq:fringe} as
\begin{equation}
  \label{eq:Noise}
  y = y_0 - \frac{C}{2}\cos(\phi + \delta\varphi) + \delta u,
\end{equation}
where $\phi$ is the total phase estimate, $\delta\varphi$ represents the phase noise, and $\delta u$ the detection noise.

\begin{figure}[!t]
  \includegraphics[width=0.48\textwidth]{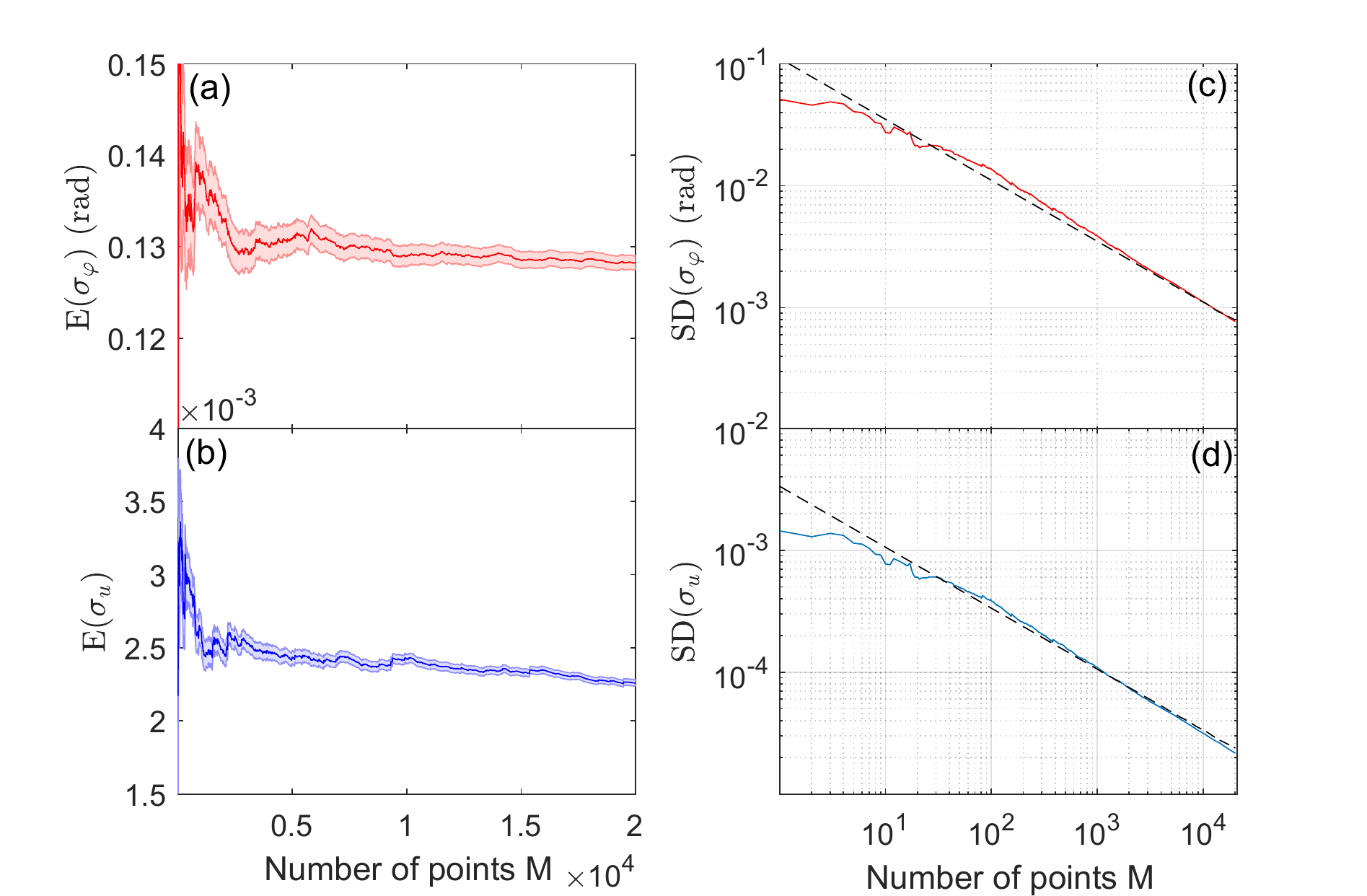}
  \caption{Expected value of the phase noise $\mathrm{E}[\sigma_\varphi]$ (a) and detection noise $\mathrm{E}[\sigma_u$] (b) as a function of the number of points $M$ considered in the dataset. After an initial transitory behavior, the estimated values converge toward their true values. (c-d) Standard deviations of the phase and detection noise estimates (also displayed as shaded areas in a,b). The standard deviations decrease as $1/\sqrt{M}$ for large number of points $M$.}
  \label{fig:NoiseEstimation}
\end{figure}

We use a Bayesian approach to estimate the statistical properties of these noise sources. Let us denote $N$ an abstract noise model, and $D$ a dataset of $M$ noisy measurements $y_i$. The probability distribution function $p(N|D)$ of the noise model $N$ given the dataset $D$ can be expressed using Bayes rule as
\begin{equation}
  \label{p(N|D)}
  p\left(N|D\right) = \frac{p\left(D|N\right) p\left(N\right)}{p\left(D\right)}.
\end{equation}
For uncorrelated noise, the probability of a given dataset knowing the noise model can be expressed as a product over the individual measurements
\begin{equation} \label{eq:product}
  p\left(D|N\right) = \prod_{i=1}^{M} p\left(y_i|N\right).
\end{equation}
Here, we consider the specific case of normally-distributed phase and detection noise, $N(\sigma_{\varphi},\sigma_u)$, with standard deviations $\sigma_\varphi$ and $\sigma_u$, respectively. The distribution of individual measurements can then be easily expressed as
\begin{equation}
 p(y_i|N) \equiv p\left(y_i|\sigma_{\varphi},\sigma_u\right) = \frac{1}{\sigma_i \sqrt{2\pi}} e^{-\frac{1}{2}(y_i/\sigma_i)^2},
\end{equation}
where
\begin{equation}
  \label{sigmai}
  \sigma_i^2 =\left( \frac{\partial y}{\partial (\delta\varphi)}\bigg {|}_{\phi_i} \right)^2 \sigma_{\varphi}^2 + \left( \frac{\partial y}{\partial (\delta u)}\bigg {|}_{\phi_i} \right)^2 \sigma_{u}^2,
\end{equation}
is the variance of the overall Gaussian noise evaluated at the phase $\phi_i$ of the $i^{\rm th}$ measurement. In the absence of prior information, the distribution $p(N) \equiv p(\sigma_{\varphi},\sigma_u)$ is chosen as uniform and $p(D)$ is simply a normalization factor. The probability distribution $p(N|D)$ can then be computed easily using \Eqs \eqref{p(N|D)} -- \eqref{sigmai}, and statistical quantities that characterize the phase and detection noise can be obtained separately by integrating this distribution
\begin{subequations}
\begin{align}
  \mathrm{E}[\sigma_\varphi]
  & = \int \sigma_\varphi p(\sigma_{\varphi},\sigma_u|D) \dd\sigma_{\varphi} \dd\sigma_u, \\
  \mathrm{E}[\sigma_u]
  & = \int \sigma_u p(\sigma_{\varphi},\sigma_u|D) \dd\sigma_{\varphi} \dd\sigma_u, \\ \mathrm{SD}[\sigma_{\varphi}]
  & = \sqrt{\mathrm{E}[{\sigma_\varphi}^2]-\mathrm{E}[{\sigma_\varphi}]^2}, \\
  \mathrm{SD}[\sigma_{u}]
  & = \sqrt{\mathrm{E}[{\sigma_u}^2] - \mathrm{E}[{\sigma_u}]^2},
\end{align}
\end{subequations}
where $\mathrm{SD}[\cdots]$ denotes the standard deviation.

Figures \ref{fig:NoiseEstimation}(a-b) show the phase and detection noise estimates calculated for a subset of the data shown in \Fig \ref{fig:StateVector}. After a brief transitory behavior, the noise estimates converge toward their true values. Figure \ref{fig:NoiseEstimation}(c-d) show the uncertainty of the noise determination that decreases as $1/\sqrt{M}$. We point out that this Bayesian method of noise characterization is an optimal and unbiased estimator, and can be applied in real time in order to adapt the KF parameters. Note also that this method can be easily generalized to non-Gaussian noise distributions or even to generic distributions \cite{Stockton2007, Barrett2015} at the expense of more computational complexity.

\subsection{Stochastic driving optimization}
\label{sec:Optimization}

\begin{figure}[!b]
  \centering
  \includegraphics[width=0.48\textwidth]{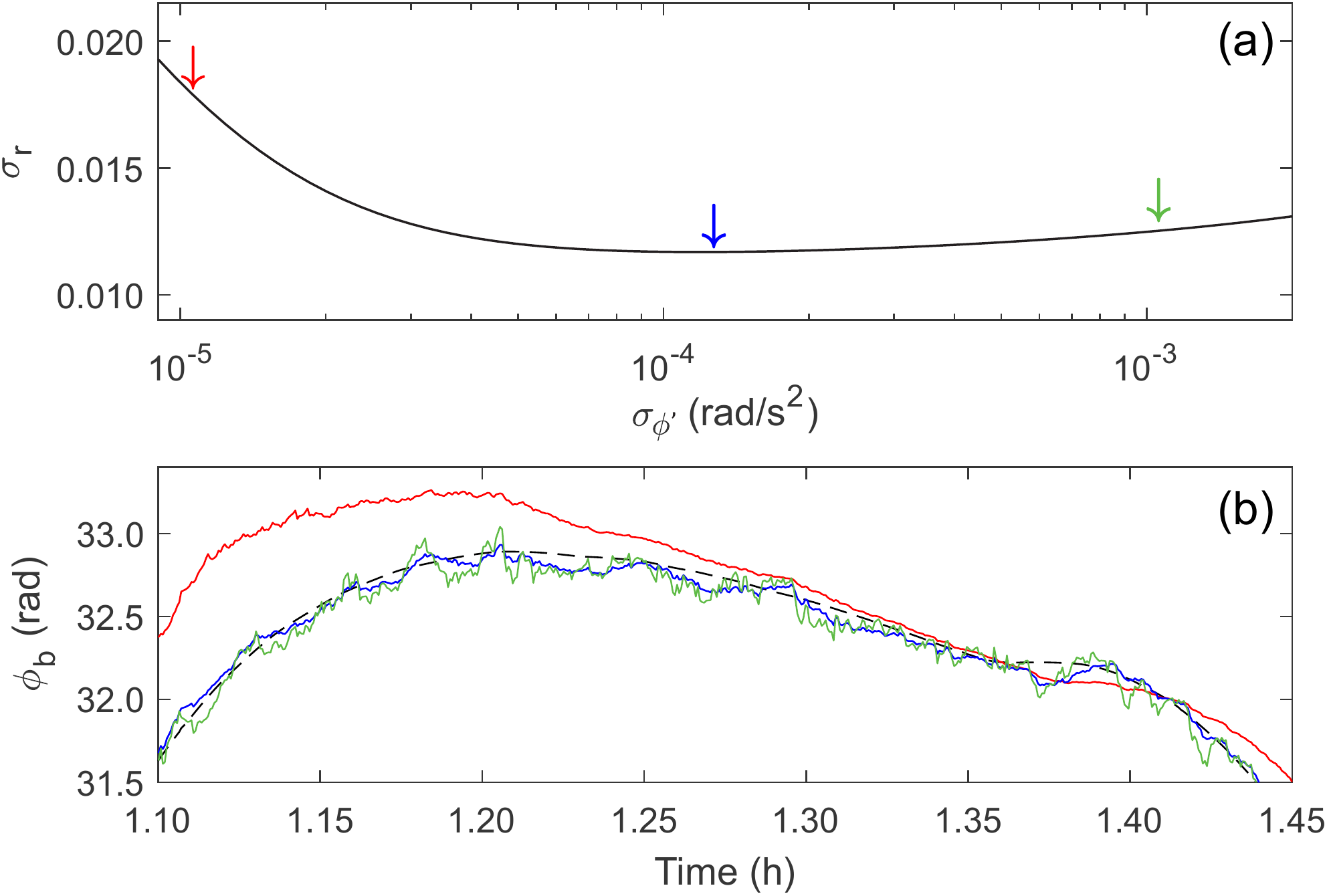}
  \caption{(a) Standard deviation of the innovation $\sigma_r$ as a function of the stochastic driving parameter $\sigma_{\phi'}$. $\sigma_r$ is minimized for $\sigma_{\phi'} = 1.2 \times 10^{-4}~\mathrm{rad/s^2}$. (b) Zoom on the bias phase tracking for $\sigma_{\phi'} = 1 \times 10^{-5}~\mathrm{rad/s^2}$ (red), $1.2 \times 10^{-4}~\mathrm{rad/s^2}$ (blue) and $1 \times 10^{-3}~\mathrm{rad/s^2}$ (green). The dashed black curve corresponds to the true bias obtained by direct averaging. The tracking lags behind for under-estimated driving and is noisier for over-estimated driving.}
  \label{fig:PhaseVarianceOptimization}
\end{figure}

Figure \ref{fig:PhaseVarianceOptimization}(a) shows the optimization of the KF over the stochastic driving variable $\sigma_{\phi'}$. $\sigma_r$ is minimized for $\sigma_{\phi'} = 1.2 \times 10^{-4}~\mathrm{rad/s^2}$. Note that the sensitivity of the innovation variance to deviations of the driving parameters from the optimum is generally very small---reflecting the robustness of the KF against errors in the parameter estimates. Figure \ref{fig:PhaseVarianceOptimization}(b) displays an example of the bias phase tracking for optimal, under- and over-estimated values of $\sigma_{\phi'}$. When the driving estimation is too small, the KF does not allow fast variations of the phase and the reconstruction lags behind. On the other hand, when the driving is too large, the KF follows too tightly the output of each measurement---adding noise to the tracked waveform.

\section{Kalman filter consistency}
\subsection{Experimental consistency checks}
\label{sec:Consistency}

A good estimator should be unbiased, and the estimate covariance should correspond to the true variance of the error. These properties constitute the consistency of the KF and ultimately depend on the true error properties. With real data, the true signal is generally not accessible, but some consistency checks can be tested on the KF innovation. Specifically, the innovation should be uncorrelated, unbiased and its probability distribution should be consistent with the probability distribution of the measurement noise. In this section, we verify the properties of the innovation on the hybrid sensor presented in \Fig \ref{fig:HybridSensor}.

Scaling the AI signal by the contrast and removing the offset, the output of the AI can be written as
\begin{equation}
  \label{eq:ScaledOutput}
  n = -\cos(\phi+\delta \varphi) + \delta u.
\end{equation}
The AI output is the combination of three independent random variables. The AI phase $\phi$ comprises the inertial and laser phases, and can be considered to be uniformly distributed over $2\pi$ for large vibration noise. The phase noise $\delta \varphi$ and detection noise $\delta u$ are normally-distributed variables with deviations $\sigma_\varphi$ and $\sigma_u$, respectively. For phase noise $\delta\varphi \ll \pi$, the error of the (scaled) innovation can be deduced from \Eq \eqref{eq:ScaledOutput} and reads
\begin{equation}
  \label{eq:ScaledNoise}
  \delta n = \sin(\phi) \delta\varphi + \delta u.
\end{equation}
Due to the sinusoidal output of the AI, even for normally-distributed phase and detection noise, the combined noise is non-Gaussian. Considering phase noise only, the probability density function of the innovation error $\delta n$ can be calculated analytically as
\begin{equation}
  \label{eq:PDFPhase}
  p(\delta n) = e^{-\delta n^2/2\sigma_\varphi^2} K_0 \left( \frac{\delta n^2}{4 \sigma_\varphi^2} \right),
\end{equation}
where $K_0$ is a modified Bessel function of the second kind. The detection noise can then be included by convolving this distribution with the corresponding normal distribution
\begin{equation}
  \label{eq:PDFAll}
  p(\delta n) = \int e^{-z^2/2\sigma_\varphi^2} K_0\left(\frac{z^2}{4\sigma_\varphi^2}\right) e^{-(z - \delta n)^2/2\sigma_u^2} \dd z.
\end{equation}

\begin{figure}[!tb]
  \includegraphics[width=0.48\textwidth]{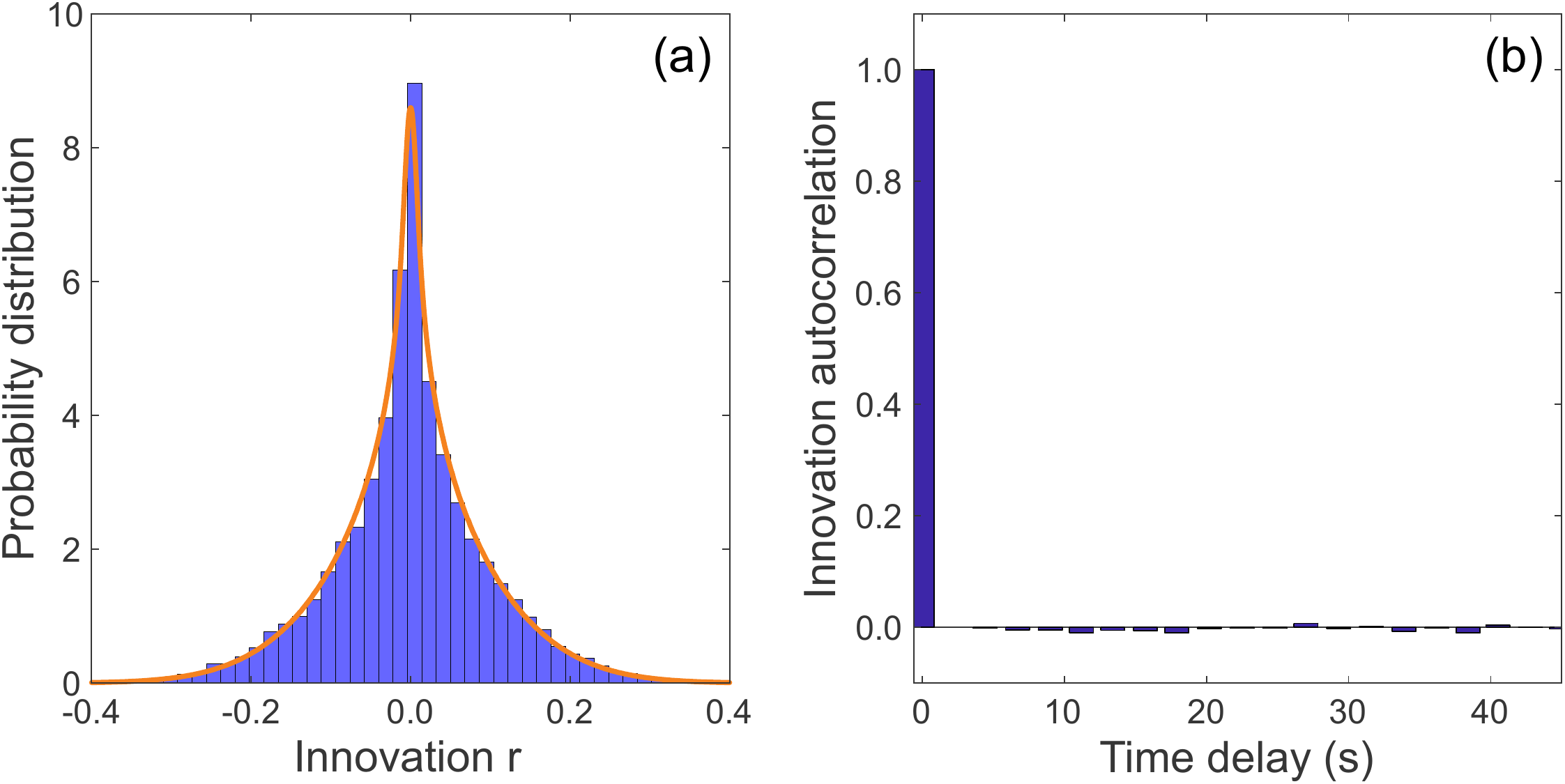}
  \caption{(a) Histogram of the innovation probability distribution, which is in excellent agreement with the analytical model of \Eq \eqref{eq:PDFAll} (orange curve). (b) Autocorrelation of the innovation as a function of delay time. After one AI cycle, no significant correlation is visible.}
  \label{fig:Consistency}
\end{figure}

Figure \ref{fig:Consistency}(a) shows a histogram of the innovation probability distribution in excellent agreement with the theoretical predictions of \Eq \eqref{eq:PDFAll}. Note that this is not an \textit{ab initio} comparison as the noise variances have been determined experimentally using the procedure described in Appendix \ref{sec:Kalman optimization}. However, it constitutes a good check of the Gaussian nature of the noise. Figure \ref{fig:Consistency}(b) shows the autocorrelation of the innovation as a function of delay time. Already after a delay of one AI cycle ($\sim 1.25$ s), no significant correlations subsist. The complete absence of correlation likely stems from an aliasing effect. Indeed, the AI interrogation time is small compared to the cycling time so that the phase estimation or detection error are not correlated between two successive AI shots. This confirms the white noise hypothesis used in the KF.

\subsection{Monte-Carlo consistency checks}

\begin{figure}[!t]
  \includegraphics[width=0.48\textwidth]{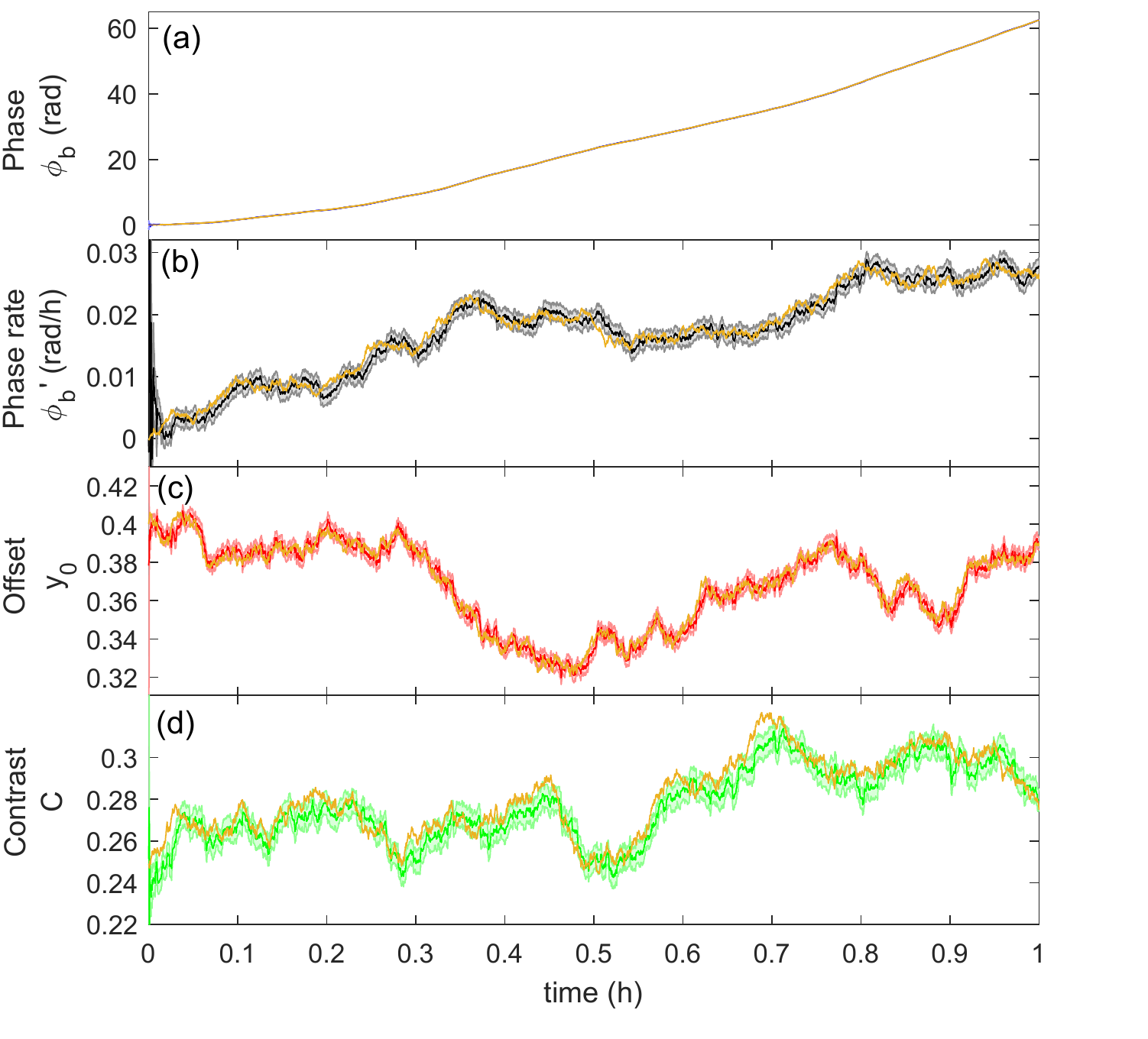}
  \caption{Tracking of the bias phase (a), bias phase rate (b), offset (c) and contrast (d) using the non-linear KF with a simulated waveform and dataset. The shaded areas correspond to the standard deviations estimated by the KF, and the yellow line is the true waveform.}
  \label{fig:SimulationTracking}
\end{figure}

To further test the consistency of our KF, we use simulated data. It is thus possible to produce waveforms that follow exactly the dynamics of the stochastic equation \eqref{Eq:stochDyn}. In addition, this permits one to compare the KF estimate to the true value of all components of the state vector (and not only the bias phase). We generate a waveform and an AI dataset that includes phase and detection noise, and we apply the KF with the true driving and noise parameters. Figure \ref{fig:SimulationTracking} shows the tracking of the state vector along with its true value for the optimal parameters obtained in our experiment. We observe that the KF tracks efficiently the waveform and that the estimation standard deviation corresponds to the typical true error.

We now compute the true estimator bias and standard deviations. To do so, we generate 1000 independent waveforms and measurement datasets to which we apply the KF, and we compare the waveform estimate to its true value. Table \ref{Table:Simulations} shows the true bias and RMS error of the estimation along with the estimation standard deviation for the bias phase, offset and contrast. We verify that the bias phase and offset estimates are unbiased and the true error RMS is in excellent agreement with the estimation standard deviation. However, the contrast estimation is slightly biased---showing that the KF tends to underestimate the contrast. This effect is already visible in \Fig \ref{fig:SimulationTracking}(d) and originates from the linear approximation that is used to compute the expected AI output. During the measurement step, the KF uses the most likely AI output to update the state vector. However, because of the non-linearity of the cosine function, the expected and most likely values do not coincide, and the KF systematically over- (under-) estimates the expected output on the top (bottom) of the fringe. The KF adapts to this error by reducing the contrast estimation. For our experimental parameters, this does not significantly affect the phase and offset estimations but it could affect the KF performances for larger phase noise. This issue can be efficiently solved by using other KF extensions such as Monte-Carlo or unscented filters which are beyond the scope of this article.

\begin{table}[!b]

  \begin{tabular}{lccc}
    \hline \hline
     & True error bias & True error RMS & Estimation SD \\ \hline
    Bias phase & $-1.1\times10^{-4}$ & $4.025\times10^{-2}$ &  $4.016\times10^{-2}$ \\
    Offset & $+5.4\times10^{-6}$ & $3.45\times10^{-3}$ & $3.43\times 10^{-3}$        \\
    Contrast & $-2.6\times10^{-3}$ &  $6.01\times10^{-3}$ & $5.34\times10^{-3}$     \\
    \hline
  \end{tabular}
  \caption{Monte-Carlo simulation results for the true error bias, true error RMS, and estimation standard deviation of each waveform parameter estimate. The bias phase is listed in radians, while the offset and contrast are unitless quantities.}
  \label{Table:Simulations}
\end{table}


\begin{thebibliography}{50}%
\makeatletter
\providecommand \@ifxundefined [1]{%
 \@ifx{#1\undefined}
}%
\providecommand \@ifnum [1]{%
 \ifnum #1\expandafter \@firstoftwo
 \else \expandafter \@secondoftwo
 \fi
}%
\providecommand \@ifx [1]{%
 \ifx #1\expandafter \@firstoftwo
 \else \expandafter \@secondoftwo
 \fi
}%
\providecommand \natexlab [1]{#1}%
\providecommand \enquote  [1]{``#1''}%
\providecommand \bibnamefont  [1]{#1}%
\providecommand \bibfnamefont [1]{#1}%
\providecommand \citenamefont [1]{#1}%
\providecommand \href@noop [0]{\@secondoftwo}%
\providecommand \href [0]{\begingroup \@sanitize@url \@href}%
\providecommand \@href[1]{\@@startlink{#1}\@@href}%
\providecommand \@@href[1]{\endgroup#1\@@endlink}%
\providecommand \@sanitize@url [0]{\catcode `\\12\catcode `\$12\catcode
  `\&12\catcode `\#12\catcode `\^12\catcode `\_12\catcode `\%12\relax}%
\providecommand \@@startlink[1]{}%
\providecommand \@@endlink[0]{}%
\providecommand \url  [0]{\begingroup\@sanitize@url \@url }%
\providecommand \@url [1]{\endgroup\@href {#1}{\urlprefix }}%
\providecommand \urlprefix  [0]{URL }%
\providecommand \Eprint [0]{\href }%
\providecommand \doibase [0]{http://dx.doi.org/}%
\providecommand \selectlanguage [0]{\@gobble}%
\providecommand \bibinfo  [0]{\@secondoftwo}%
\providecommand \bibfield  [0]{\@secondoftwo}%
\providecommand \translation [1]{[#1]}%
\providecommand \BibitemOpen [0]{}%
\providecommand \bibitemStop [0]{}%
\providecommand \bibitemNoStop [0]{.\EOS\space}%
\providecommand \EOS [0]{\spacefactor3000\relax}%
\providecommand \BibitemShut  [1]{\csname bibitem#1\endcsname}%
\let\auto@bib@innerbib\@empty
\bibitem [{\citenamefont {Titterton}\ and\ \citenamefont
  {Weston}(2004)}]{Titterton2004}%
  \BibitemOpen
  \bibfield  {author} {\bibinfo {author} {\bibfnamefont {D.~H.}\ \bibnamefont
  {Titterton}}\ and\ \bibinfo {author} {\bibfnamefont {J.~L.}\ \bibnamefont
  {Weston}},\ }\href {\doibase 10.1049/PBRA017E} {\emph {\bibinfo {title}
  {{Strapdown Inertial Navigation Technology}}}},\ \bibinfo {edition} {2nd}\
  ed.,\ Electromagnetics and Radar Series\ (\bibinfo  {publisher} {Institution
  of Engineering and Technology},\ \bibinfo {address} {London, UK},\ \bibinfo
  {year} {2004})\BibitemShut {NoStop}%
\bibitem [{\citenamefont {Schuler}(1923)}]{Schuler1923}%
  \BibitemOpen
  \bibfield  {author} {\bibinfo {author} {\bibfnamefont {M.}~\bibnamefont
  {Schuler}},\ }\bibfield  {title} {\enquote {\bibinfo {title} {{The
  Perturbation of Pendulum and Gyroscope Instruments by Acceleration of the
  Vehicule}},}\ }\href {http://www.dtic.mil/dtic/tr/fulltext/u2/b806120.pdf}
  {\bibfield  {journal} {\bibinfo  {journal} {Physik. Zeitschr.}\ }\textbf
  {\bibinfo {volume} {24}},\ \bibinfo {pages} {344} (\bibinfo {year}
  {1923})}\BibitemShut {NoStop}%
\bibitem [{\citenamefont {Bouchendira}\ \emph {et~al.}(2011)\citenamefont
  {Bouchendira}, \citenamefont {Clad\'{e}}, \citenamefont
  {Guellati-Kh{\'{e}}lifa}, \citenamefont {Nez},\ and\ \citenamefont
  {Biraben}}]{Bouchendira2011}%
  \BibitemOpen
  \bibfield  {author} {\bibinfo {author} {\bibfnamefont {R.}~\bibnamefont
  {Bouchendira}}, \bibinfo {author} {\bibfnamefont {P.}~\bibnamefont
  {Clad\'{e}}}, \bibinfo {author} {\bibfnamefont {S.}~\bibnamefont
  {Guellati-Kh{\'{e}}lifa}}, \bibinfo {author} {\bibfnamefont {F.}~\bibnamefont
  {Nez}}, \ and\ \bibinfo {author} {\bibfnamefont {F.}~\bibnamefont
  {Biraben}},\ }\bibfield  {title} {\enquote {\bibinfo {title} {{New
  Determination of the Fine Structure Constant and Test of the Quantum
  Electrodynamics}},}\ }\href {\doibase 10.1103/PhysRevLett.106.080801}
  {\bibfield  {journal} {\bibinfo  {journal} {Phys. Rev. Lett.}\ }\textbf
  {\bibinfo {volume} {106}},\ \bibinfo {pages} {080801} (\bibinfo {year}
  {2011})}\BibitemShut {NoStop}%
\bibitem [{\citenamefont {Gregoire}\ \emph {et~al.}(2015)\citenamefont
  {Gregoire}, \citenamefont {Hromada}, \citenamefont {Holmgren}, \citenamefont
  {Trubko},\ and\ \citenamefont {Cronin}}]{Cronin2015}%
  \BibitemOpen
  \bibfield  {author} {\bibinfo {author} {\bibfnamefont {M.~D.}\ \bibnamefont
  {Gregoire}}, \bibinfo {author} {\bibfnamefont {I.}~\bibnamefont {Hromada}},
  \bibinfo {author} {\bibfnamefont {W.~F.}\ \bibnamefont {Holmgren}}, \bibinfo
  {author} {\bibfnamefont {R.}~\bibnamefont {Trubko}}, \ and\ \bibinfo {author}
  {\bibfnamefont {A.~D.}\ \bibnamefont {Cronin}},\ }\bibfield  {title}
  {\enquote {\bibinfo {title} {Measurements of the ground-state
  polarizabilities of {C}s, {R}b, and {K} using atom interferometry},}\ }\href
  {\doibase 10.1103/PhysRevA.92.052513} {\bibfield  {journal} {\bibinfo
  {journal} {Phys. Rev. A}\ }\textbf {\bibinfo {volume} {92}},\ \bibinfo
  {pages} {052513} (\bibinfo {year} {2015})}\BibitemShut {NoStop}%
\bibitem [{\citenamefont {Zhou}\ \emph {et~al.}(2015)\citenamefont {Zhou},
  \citenamefont {Long}, \citenamefont {Tang}, \citenamefont {Chen},
  \citenamefont {Gao}, \citenamefont {Peng}, \citenamefont {Duan},
  \citenamefont {Zhong}, \citenamefont {Xiong}, \citenamefont {Wang},
  \citenamefont {Zhang},\ and\ \citenamefont {Zhan}}]{LZhou2015}%
  \BibitemOpen
  \bibfield  {author} {\bibinfo {author} {\bibfnamefont {L.}~\bibnamefont
  {Zhou}}, \bibinfo {author} {\bibfnamefont {S.}~\bibnamefont {Long}}, \bibinfo
  {author} {\bibfnamefont {B.}~\bibnamefont {Tang}}, \bibinfo {author}
  {\bibfnamefont {X.}~\bibnamefont {Chen}}, \bibinfo {author} {\bibfnamefont
  {F.}~\bibnamefont {Gao}}, \bibinfo {author} {\bibfnamefont {W.}~\bibnamefont
  {Peng}}, \bibinfo {author} {\bibfnamefont {We.}\ \bibnamefont {Duan}},
  \bibinfo {author} {\bibfnamefont {J.}~\bibnamefont {Zhong}}, \bibinfo
  {author} {\bibfnamefont {Z.}~\bibnamefont {Xiong}}, \bibinfo {author}
  {\bibfnamefont {J.}~\bibnamefont {Wang}}, \bibinfo {author} {\bibfnamefont
  {Y.}~\bibnamefont {Zhang}}, \ and\ \bibinfo {author} {\bibfnamefont
  {M.}~\bibnamefont {Zhan}},\ }\bibfield  {title} {\enquote {\bibinfo {title}
  {{Test of Equivalence Principle at $10^{-8}$ Level by a Dual-Species
  Double-Diffraction Raman Atom Interferometer}},}\ }\href {\doibase
  10.1103/PhysRevLett.115.013004} {\bibfield  {journal} {\bibinfo  {journal}
  {Phys. Rev. Lett.}\ }\textbf {\bibinfo {volume} {115}},\ \bibinfo {pages}
  {013004} (\bibinfo {year} {2015})}\BibitemShut {NoStop}%
\bibitem [{\citenamefont {Kovachy}\ \emph {et~al.}(2015)\citenamefont
  {Kovachy}, \citenamefont {Asenbaum}, \citenamefont {Overstreet},
  \citenamefont {Donnelly}, \citenamefont {Dickerson}, \citenamefont
  {Sugarbaker}, \citenamefont {Hogan},\ and\ \citenamefont
  {Kasevich}}]{Kovachy2015}%
  \BibitemOpen
  \bibfield  {author} {\bibinfo {author} {\bibfnamefont {T.}~\bibnamefont
  {Kovachy}}, \bibinfo {author} {\bibfnamefont {P.}~\bibnamefont {Asenbaum}},
  \bibinfo {author} {\bibfnamefont {C.}~\bibnamefont {Overstreet}}, \bibinfo
  {author} {\bibfnamefont {C.~A.}\ \bibnamefont {Donnelly}}, \bibinfo {author}
  {\bibfnamefont {S.~M.}\ \bibnamefont {Dickerson}}, \bibinfo {author}
  {\bibfnamefont {A.}~\bibnamefont {Sugarbaker}}, \bibinfo {author}
  {\bibfnamefont {J.~M.}\ \bibnamefont {Hogan}}, \ and\ \bibinfo {author}
  {\bibfnamefont {M.~A.}\ \bibnamefont {Kasevich}},\ }\bibfield  {title}
  {\enquote {\bibinfo {title} {{Quantum superposition at the half-metre
  scale}},}\ }\href {\doibase 10.1038/nature16155} {\bibfield  {journal}
  {\bibinfo  {journal} {Nature}\ }\textbf {\bibinfo {volume} {528}},\ \bibinfo
  {pages} {530--533} (\bibinfo {year} {2015})}\BibitemShut {NoStop}%
\bibitem [{\citenamefont {Barrett}\ \emph
  {et~al.}(2016{\natexlab{a}})\citenamefont {Barrett}, \citenamefont {Carew},
  \citenamefont {Beica}, \citenamefont {Vorozcovs}, \citenamefont {Pouliot},\
  and\ \citenamefont {Kumarakrishnan}}]{Barrett2016b}%
  \BibitemOpen
  \bibfield  {author} {\bibinfo {author} {\bibfnamefont {B.}~\bibnamefont
  {Barrett}}, \bibinfo {author} {\bibfnamefont {A.}~\bibnamefont {Carew}},
  \bibinfo {author} {\bibfnamefont {H.~C.}\ \bibnamefont {Beica}}, \bibinfo
  {author} {\bibfnamefont {A.}~\bibnamefont {Vorozcovs}}, \bibinfo {author}
  {\bibfnamefont {A.}~\bibnamefont {Pouliot}}, \ and\ \bibinfo {author}
  {\bibfnamefont {A.}~\bibnamefont {Kumarakrishnan}},\ }\bibfield  {title}
  {\enquote {\bibinfo {title} {{Prospects for Precise Measurements with Echo
  Atom Interferometry}},}\ }\href {\doibase 10.3390/atoms4030019} {\bibfield
  {journal} {\bibinfo  {journal} {Atoms}\ }\textbf {\bibinfo {volume} {4}},\
  \bibinfo {pages} {19} (\bibinfo {year} {2016}{\natexlab{a}})}\BibitemShut
  {NoStop}%
\bibitem [{\citenamefont {Rosi}\ \emph {et~al.}(2017)\citenamefont {Rosi},
  \citenamefont {D'Amico}, \citenamefont {Cacciapuoti}, \citenamefont
  {Sorrentino}, \citenamefont {Prevedelli}, \citenamefont {Zych}, \citenamefont
  {Brukner},\ and\ \citenamefont {Tino}}]{Rosi2017}%
  \BibitemOpen
  \bibfield  {author} {\bibinfo {author} {\bibfnamefont {G.}~\bibnamefont
  {Rosi}}, \bibinfo {author} {\bibfnamefont {G.}~\bibnamefont {D'Amico}},
  \bibinfo {author} {\bibfnamefont {L.}~\bibnamefont {Cacciapuoti}}, \bibinfo
  {author} {\bibfnamefont {F.}~\bibnamefont {Sorrentino}}, \bibinfo {author}
  {\bibfnamefont {M.}~\bibnamefont {Prevedelli}}, \bibinfo {author}
  {\bibfnamefont {M.}~\bibnamefont {Zych}}, \bibinfo {author} {\bibfnamefont
  {C.}~\bibnamefont {Brukner}}, \ and\ \bibinfo {author} {\bibfnamefont
  {G.~M.}\ \bibnamefont {Tino}},\ }\bibfield  {title} {\enquote {\bibinfo
  {title} {{Quantum test of the equivalence principle for atoms in coherent
  superposition of internal energy states}},}\ }\href {\doibase
  10.1038/ncomms15529} {\bibfield  {journal} {\bibinfo  {journal} {Nat.
  Commun.}\ }\textbf {\bibinfo {volume} {8}},\ \bibinfo {pages} {15529}
  (\bibinfo {year} {2017})}\BibitemShut {NoStop}%
\bibitem [{\citenamefont {Peters}\ \emph {et~al.}(2001)\citenamefont {Peters},
  \citenamefont {Chung},\ and\ \citenamefont {Chu}}]{Peters2001}%
  \BibitemOpen
  \bibfield  {author} {\bibinfo {author} {\bibfnamefont {A.}~\bibnamefont
  {Peters}}, \bibinfo {author} {\bibfnamefont {K.~Y.}\ \bibnamefont {Chung}}, \
  and\ \bibinfo {author} {\bibfnamefont {S.}~\bibnamefont {Chu}},\ }\bibfield
  {title} {\enquote {\bibinfo {title} {{High-precision gravity measurements
  using atom interferometry}},}\ }\href
  {http://stacks.iop.org/0026-1394/38/i=1/a=4} {\bibfield  {journal} {\bibinfo
  {journal} {Metrologia}\ }\textbf {\bibinfo {volume} {38}},\ \bibinfo {pages}
  {25} (\bibinfo {year} {2001})}\BibitemShut {NoStop}%
\bibitem [{\citenamefont {Le~Gou{\"e}t}\ \emph {et~al.}(2008)\citenamefont
  {Le~Gou{\"e}t}, \citenamefont {Mehlst{\"a}ubler}, \citenamefont {Kim},
  \citenamefont {Merlet}, \citenamefont {Clairon}, \citenamefont {Landragin},\
  and\ \citenamefont {{Pereira Dos Santos}}}]{LeGouet2008}%
  \BibitemOpen
  \bibfield  {author} {\bibinfo {author} {\bibfnamefont {J.}~\bibnamefont
  {Le~Gou{\"e}t}}, \bibinfo {author} {\bibfnamefont {T.~E.}\ \bibnamefont
  {Mehlst{\"a}ubler}}, \bibinfo {author} {\bibfnamefont {J.}~\bibnamefont
  {Kim}}, \bibinfo {author} {\bibfnamefont {S.}~\bibnamefont {Merlet}},
  \bibinfo {author} {\bibfnamefont {A.}~\bibnamefont {Clairon}}, \bibinfo
  {author} {\bibfnamefont {A.}~\bibnamefont {Landragin}}, \ and\ \bibinfo
  {author} {\bibfnamefont {F.}~\bibnamefont {{Pereira Dos Santos}}},\
  }\bibfield  {title} {\enquote {\bibinfo {title} {{Limits to the sensitivity
  of a low noise compact atomic gravimeter}},}\ }\href {\doibase
  10.1007/s00340-008-3088-1} {\bibfield  {journal} {\bibinfo  {journal} {Appl.
  Phys. B}\ }\textbf {\bibinfo {volume} {92}},\ \bibinfo {pages} {133--144}
  (\bibinfo {year} {2008})}\BibitemShut {NoStop}%
\bibitem [{\citenamefont {Altin}\ \emph {et~al.}(2013)\citenamefont {Altin},
  \citenamefont {Johnsson}, \citenamefont {Dennis}, \citenamefont {Anderson},
  \citenamefont {Debs}, \citenamefont {Szigeti}, \citenamefont {Hardman},
  \citenamefont {Bennetts}, \citenamefont {McDonald}, \citenamefont {Turner},
  \citenamefont {Close}, \citenamefont {Robins},\ and\ \citenamefont
  {Negnevitsky}}]{Altin2013}%
  \BibitemOpen
  \bibfield  {author} {\bibinfo {author} {\bibfnamefont {P.~A.}\ \bibnamefont
  {Altin}}, \bibinfo {author} {\bibfnamefont {M.~T.}\ \bibnamefont {Johnsson}},
  \bibinfo {author} {\bibfnamefont {G.~R.}\ \bibnamefont {Dennis}}, \bibinfo
  {author} {\bibfnamefont {R.~P.}\ \bibnamefont {Anderson}}, \bibinfo {author}
  {\bibfnamefont {J.~E.}\ \bibnamefont {Debs}}, \bibinfo {author}
  {\bibfnamefont {S.~S.}\ \bibnamefont {Szigeti}}, \bibinfo {author}
  {\bibfnamefont {K.~S.}\ \bibnamefont {Hardman}}, \bibinfo {author}
  {\bibfnamefont {S.}~\bibnamefont {Bennetts}}, \bibinfo {author}
  {\bibfnamefont {G.~D.}\ \bibnamefont {McDonald}}, \bibinfo {author}
  {\bibfnamefont {L.~D.}\ \bibnamefont {Turner}}, \bibinfo {author}
  {\bibfnamefont {J.~D.}\ \bibnamefont {Close}}, \bibinfo {author}
  {\bibfnamefont {N.~P.}\ \bibnamefont {Robins}}, \ and\ \bibinfo {author}
  {\bibfnamefont {V.}~\bibnamefont {Negnevitsky}},\ }\bibfield  {title}
  {\enquote {\bibinfo {title} {{Precision atomic gravimeter based on Bragg
  diffraction}},}\ }\href {\doibase 10.1088/1367-2630/15/2/023009} {\bibfield
  {journal} {\bibinfo  {journal} {New J. Phys.}\ }\textbf {\bibinfo {volume}
  {15}},\ \bibinfo {pages} {023009} (\bibinfo {year} {2013})}\BibitemShut
  {NoStop}%
\bibitem [{\citenamefont {Gillot}\ \emph {et~al.}(2014)\citenamefont {Gillot},
  \citenamefont {Francis}, \citenamefont {Landragin}, \citenamefont {{Pereira
  Dos Santos}},\ and\ \citenamefont {Merlet}}]{Gillot2014}%
  \BibitemOpen
  \bibfield  {author} {\bibinfo {author} {\bibfnamefont {P.}~\bibnamefont
  {Gillot}}, \bibinfo {author} {\bibfnamefont {O.}~\bibnamefont {Francis}},
  \bibinfo {author} {\bibfnamefont {A.}~\bibnamefont {Landragin}}, \bibinfo
  {author} {\bibfnamefont {F.}~\bibnamefont {{Pereira Dos Santos}}}, \ and\
  \bibinfo {author} {\bibfnamefont {S.}~\bibnamefont {Merlet}},\ }\bibfield
  {title} {\enquote {\bibinfo {title} {{Stability comparison of two absolute
  gravimeters: optical versus atomic interferometers}},}\ }\href {\doibase
  10.1088/0026-1394/51/5/l15} {\bibfield  {journal} {\bibinfo  {journal}
  {Metrologia}\ }\textbf {\bibinfo {volume} {51}},\ \bibinfo {pages} {L15--L17}
  (\bibinfo {year} {2014})}\BibitemShut {NoStop}%
\bibitem [{\citenamefont {Freier}\ \emph {et~al.}(2016)\citenamefont {Freier},
  \citenamefont {Hauth}, \citenamefont {Schkolnik}, \citenamefont {Leykauf},
  \citenamefont {Schilling}, \citenamefont {Wziontek}, \citenamefont
  {Scherneck}, \citenamefont {Muller},\ and\ \citenamefont
  {Peters}}]{Freier2016}%
  \BibitemOpen
  \bibfield  {author} {\bibinfo {author} {\bibfnamefont {C.}~\bibnamefont
  {Freier}}, \bibinfo {author} {\bibfnamefont {M.}~\bibnamefont {Hauth}},
  \bibinfo {author} {\bibfnamefont {V.}~\bibnamefont {Schkolnik}}, \bibinfo
  {author} {\bibfnamefont {B.}~\bibnamefont {Leykauf}}, \bibinfo {author}
  {\bibfnamefont {M.}~\bibnamefont {Schilling}}, \bibinfo {author}
  {\bibfnamefont {H.}~\bibnamefont {Wziontek}}, \bibinfo {author}
  {\bibfnamefont {H.-G.}\ \bibnamefont {Scherneck}}, \bibinfo {author}
  {\bibfnamefont {J.}~\bibnamefont {Muller}}, \ and\ \bibinfo {author}
  {\bibfnamefont {A.}~\bibnamefont {Peters}},\ }\bibfield  {title} {\enquote
  {\bibinfo {title} {{Mobile quantum gravity sensor with unprecedented
  stability}},}\ }\href {\doibase 10.1088/1742-6596/723/1/012050} {\bibfield
  {journal} {\bibinfo  {journal} {J. Phys. Conf. Ser.}\ }\textbf {\bibinfo
  {volume} {723}},\ \bibinfo {pages} {012050} (\bibinfo {year}
  {2016})}\BibitemShut {NoStop}%
\bibitem [{\citenamefont {Hardman}\ \emph {et~al.}(2016)\citenamefont
  {Hardman}, \citenamefont {Everitt}, \citenamefont {McDonald}, \citenamefont
  {Manju}, \citenamefont {Wigley}, \citenamefont {Sooriyabandara},
  \citenamefont {Kuhn}, \citenamefont {Debs}, \citenamefont {Close},\ and\
  \citenamefont {Robins}}]{Hardman2016}%
  \BibitemOpen
  \bibfield  {author} {\bibinfo {author} {\bibfnamefont {K.~S.}\ \bibnamefont
  {Hardman}}, \bibinfo {author} {\bibfnamefont {P.~J.}\ \bibnamefont
  {Everitt}}, \bibinfo {author} {\bibfnamefont {G.~D.}\ \bibnamefont
  {McDonald}}, \bibinfo {author} {\bibfnamefont {P.}~\bibnamefont {Manju}},
  \bibinfo {author} {\bibfnamefont {P.~B.}\ \bibnamefont {Wigley}}, \bibinfo
  {author} {\bibfnamefont {M.~A.}\ \bibnamefont {Sooriyabandara}}, \bibinfo
  {author} {\bibfnamefont {C.~C.~N.}\ \bibnamefont {Kuhn}}, \bibinfo {author}
  {\bibfnamefont {J.~E.}\ \bibnamefont {Debs}}, \bibinfo {author}
  {\bibfnamefont {J.~D.}\ \bibnamefont {Close}}, \ and\ \bibinfo {author}
  {\bibfnamefont {N.~P.}\ \bibnamefont {Robins}},\ }\bibfield  {title}
  {\enquote {\bibinfo {title} {{Simultaneous Precision Gravimetry and Magnetic
  Gradiometry with a Bose-Einstein Condensate: A High Precision, Quantum
  Sensor}},}\ }\href {\doibase 10.1103/PhysRevLett.117.138501} {\bibfield
  {journal} {\bibinfo  {journal} {Phys. Rev. Lett.}\ }\textbf {\bibinfo
  {volume} {117}},\ \bibinfo {pages} {138501} (\bibinfo {year}
  {2016})}\BibitemShut {NoStop}%
\bibitem [{\citenamefont {Bodart}\ \emph {et~al.}(2010)\citenamefont {Bodart},
  \citenamefont {Merlet}, \citenamefont {Malossi}, \citenamefont {{Pereira Dos
  Santos}}, \citenamefont {Bouyer},\ and\ \citenamefont
  {Landragin}}]{Bodart2010}%
  \BibitemOpen
  \bibfield  {author} {\bibinfo {author} {\bibfnamefont {Q.}~\bibnamefont
  {Bodart}}, \bibinfo {author} {\bibfnamefont {S.}~\bibnamefont {Merlet}},
  \bibinfo {author} {\bibfnamefont {N.}~\bibnamefont {Malossi}}, \bibinfo
  {author} {\bibfnamefont {F.}~\bibnamefont {{Pereira Dos Santos}}}, \bibinfo
  {author} {\bibfnamefont {P.}~\bibnamefont {Bouyer}}, \ and\ \bibinfo {author}
  {\bibfnamefont {A.}~\bibnamefont {Landragin}},\ }\bibfield  {title} {\enquote
  {\bibinfo {title} {{A cold atom pyramidal gravimeter with a single laser
  beam}},}\ }\href {\doibase 10.1063/1.3373917} {\bibfield  {journal} {\bibinfo
   {journal} {Appl. Phys. Lett.}\ }\textbf {\bibinfo {volume} {96}},\ \bibinfo
  {pages} {134101} (\bibinfo {year} {2010})}\BibitemShut {NoStop}%
\bibitem [{\citenamefont {Barrett}\ \emph
  {et~al.}(2016{\natexlab{b}})\citenamefont {Barrett}, \citenamefont
  {Bertoldi},\ and\ \citenamefont {Bouyer}}]{Barrett2016c}%
  \BibitemOpen
  \bibfield  {author} {\bibinfo {author} {\bibfnamefont {B.}~\bibnamefont
  {Barrett}}, \bibinfo {author} {\bibfnamefont {A.}~\bibnamefont {Bertoldi}}, \
  and\ \bibinfo {author} {\bibfnamefont {P.}~\bibnamefont {Bouyer}},\
  }\bibfield  {title} {\enquote {\bibinfo {title} {{Inertial quantum sensors
  using light and matter}},}\ }\href {\doibase 10.1088/0031-8949/91/5/053006}
  {\bibfield  {journal} {\bibinfo  {journal} {Phys. Scr.}\ }\textbf {\bibinfo
  {volume} {91}},\ \bibinfo {pages} {053006} (\bibinfo {year}
  {2016}{\natexlab{b}})}\BibitemShut {NoStop}%
\bibitem [{\citenamefont {Bidel}\ \emph {et~al.}(2018)\citenamefont {Bidel},
  \citenamefont {Zahzam}, \citenamefont {Blanchard}, \citenamefont {Bonnin},
  \citenamefont {Cadoret}, \citenamefont {Bresson}, \citenamefont {Rouxel},\
  and\ \citenamefont {Lequentrec-Lalancette}}]{Bidel2018}%
  \BibitemOpen
  \bibfield  {author} {\bibinfo {author} {\bibfnamefont {Y.}~\bibnamefont
  {Bidel}}, \bibinfo {author} {\bibfnamefont {N.}~\bibnamefont {Zahzam}},
  \bibinfo {author} {\bibfnamefont {C.}~\bibnamefont {Blanchard}}, \bibinfo
  {author} {\bibfnamefont {A.}~\bibnamefont {Bonnin}}, \bibinfo {author}
  {\bibfnamefont {M.}~\bibnamefont {Cadoret}}, \bibinfo {author} {\bibfnamefont
  {A.}~\bibnamefont {Bresson}}, \bibinfo {author} {\bibfnamefont
  {D.}~\bibnamefont {Rouxel}}, \ and\ \bibinfo {author} {\bibfnamefont {M.~F.}\
  \bibnamefont {Lequentrec-Lalancette}},\ }\bibfield  {title} {\enquote
  {\bibinfo {title} {{Absolute marine gravimetry with matter-wave
  interferometry}},}\ }\href {\doibase 10.1038/s41467-018-03040-2} {\bibfield
  {journal} {\bibinfo  {journal} {Nat. Commun.}\ }\textbf {\bibinfo {volume}
  {9}},\ \bibinfo {pages} {627} (\bibinfo {year} {2018})}\BibitemShut {NoStop}%
\bibitem [{AOs()}]{AOsense}%
  \BibitemOpen
  \href@noop {} {\enquote {\bibinfo {title} {Aosense website},}\ }\bibinfo
  {howpublished} {\url{www.aosense.com}},\ \bibinfo {note} {accessed: March
  2018}\BibitemShut {NoStop}%
\bibitem [{muq()}]{muquans}%
  \BibitemOpen
  \href@noop {} {\enquote {\bibinfo {title} {Muquans website},}\ }\bibinfo
  {howpublished} {\url{www.muquans.com}},\ \bibinfo {note} {accessed: March
  2018}\BibitemShut {NoStop}%
\bibitem [{\citenamefont {Jekeli}(2005)}]{Jekeli2005}%
  \BibitemOpen
  \bibfield  {author} {\bibinfo {author} {\bibfnamefont {C.}~\bibnamefont
  {Jekeli}},\ }\bibfield  {title} {\enquote {\bibinfo {title} {{Navigation
  Error Analysis of Atom Interferometer Inertial Sensor}},}\ }\href {\doibase
  10.1002/j.2161-4296.2005.tb01726.x} {\bibfield  {journal} {\bibinfo
  {journal} {Navigation}\ }\textbf {\bibinfo {volume} {52}},\ \bibinfo {pages}
  {1--14} (\bibinfo {year} {2005})}\BibitemShut {NoStop}%
\bibitem [{\citenamefont {Geiger}\ \emph {et~al.}(2011)\citenamefont {Geiger},
  \citenamefont {M\'{e}noret}, \citenamefont {Stern}, \citenamefont {Zahzam},
  \citenamefont {Cheinet}, \citenamefont {Battelier}, \citenamefont {Villing},
  \citenamefont {Moron}, \citenamefont {Lours}, \citenamefont {Bidel},
  \citenamefont {Bresson}, \citenamefont {Landragin},\ and\ \citenamefont
  {Bouyer}}]{Geiger2011}%
  \BibitemOpen
  \bibfield  {author} {\bibinfo {author} {\bibfnamefont {R.}~\bibnamefont
  {Geiger}}, \bibinfo {author} {\bibfnamefont {V.}~\bibnamefont {M\'{e}noret}},
  \bibinfo {author} {\bibfnamefont {G.}~\bibnamefont {Stern}}, \bibinfo
  {author} {\bibfnamefont {N.}~\bibnamefont {Zahzam}}, \bibinfo {author}
  {\bibfnamefont {P.}~\bibnamefont {Cheinet}}, \bibinfo {author} {\bibfnamefont
  {B.}~\bibnamefont {Battelier}}, \bibinfo {author} {\bibfnamefont
  {A.}~\bibnamefont {Villing}}, \bibinfo {author} {\bibfnamefont
  {F.}~\bibnamefont {Moron}}, \bibinfo {author} {\bibfnamefont
  {M.}~\bibnamefont {Lours}}, \bibinfo {author} {\bibfnamefont
  {Y.}~\bibnamefont {Bidel}}, \bibinfo {author} {\bibfnamefont
  {A.}~\bibnamefont {Bresson}}, \bibinfo {author} {\bibfnamefont
  {A.}~\bibnamefont {Landragin}}, \ and\ \bibinfo {author} {\bibfnamefont
  {P.}~\bibnamefont {Bouyer}},\ }\bibfield  {title} {\enquote {\bibinfo {title}
  {{Detecting inertial effects with airborne matter-wave interferometry}},}\
  }\href {\doibase 10.1038/ncomms1479} {\bibfield  {journal} {\bibinfo
  {journal} {Nat. Commun.}\ }\textbf {\bibinfo {volume} {2}},\ \bibinfo {pages}
  {474} (\bibinfo {year} {2011})}\BibitemShut {NoStop}%
\bibitem [{\citenamefont {Barrett}\ \emph
  {et~al.}(2016{\natexlab{c}})\citenamefont {Barrett}, \citenamefont
  {Antoni-Micollier}, \citenamefont {Chichet}, \citenamefont {Battelier},
  \citenamefont {L{\'{e}}v{\`{e}}que}, \citenamefont {Landragin},\ and\
  \citenamefont {Bouyer}}]{Barrett2016a}%
  \BibitemOpen
  \bibfield  {author} {\bibinfo {author} {\bibfnamefont {B.}~\bibnamefont
  {Barrett}}, \bibinfo {author} {\bibfnamefont {L.}~\bibnamefont
  {Antoni-Micollier}}, \bibinfo {author} {\bibfnamefont {L.}~\bibnamefont
  {Chichet}}, \bibinfo {author} {\bibfnamefont {B.}~\bibnamefont {Battelier}},
  \bibinfo {author} {\bibfnamefont {T.}~\bibnamefont {L{\'{e}}v{\`{e}}que}},
  \bibinfo {author} {\bibfnamefont {A.}~\bibnamefont {Landragin}}, \ and\
  \bibinfo {author} {\bibfnamefont {P.}~\bibnamefont {Bouyer}},\ }\bibfield
  {title} {\enquote {\bibinfo {title} {{Dual matter-wave inertial sensors in
  weightlessness}},}\ }\href {\doibase 10.1038/ncomms13786} {\bibfield
  {journal} {\bibinfo  {journal} {Nat. Commun.}\ }\textbf {\bibinfo {volume}
  {7}},\ \bibinfo {pages} {13786} (\bibinfo {year}
  {2016}{\natexlab{c}})}\BibitemShut {NoStop}%
\bibitem [{\citenamefont {Battelier}\ \emph {et~al.}(2016)\citenamefont
  {Battelier}, \citenamefont {Barrett}, \citenamefont {Fouch{\'{e}}},
  \citenamefont {Chichet}, \citenamefont {Antoni-Micollier}, \citenamefont
  {Porte}, \citenamefont {Napolitano}, \citenamefont {Lautier}, \citenamefont
  {Landragin},\ and\ \citenamefont {Bouyer}}]{Battelier2016}%
  \BibitemOpen
  \bibfield  {author} {\bibinfo {author} {\bibfnamefont {B.}~\bibnamefont
  {Battelier}}, \bibinfo {author} {\bibfnamefont {B.}~\bibnamefont {Barrett}},
  \bibinfo {author} {\bibfnamefont {L.}~\bibnamefont {Fouch{\'{e}}}}, \bibinfo
  {author} {\bibfnamefont {L.}~\bibnamefont {Chichet}}, \bibinfo {author}
  {\bibfnamefont {L.}~\bibnamefont {Antoni-Micollier}}, \bibinfo {author}
  {\bibfnamefont {H.}~\bibnamefont {Porte}}, \bibinfo {author} {\bibfnamefont
  {F.}~\bibnamefont {Napolitano}}, \bibinfo {author} {\bibfnamefont
  {J.}~\bibnamefont {Lautier}}, \bibinfo {author} {\bibfnamefont
  {A.}~\bibnamefont {Landragin}}, \ and\ \bibinfo {author} {\bibfnamefont
  {P.}~\bibnamefont {Bouyer}},\ }\bibfield  {title} {\enquote {\bibinfo {title}
  {{Development of compact cold-atom sensors for inertial navigation}},}\ }in\
  \href {\doibase 10.1117/12.2228351} {\emph {\bibinfo {booktitle} {Proceedings
  of SPIE Quantum Optics}}},\ Vol.\ \bibinfo {volume} {9900}\ (\bibinfo {year}
  {2016})\ p.\ \bibinfo {pages} {990004}\BibitemShut {NoStop}%
\bibitem [{\citenamefont {Rakholia}\ \emph {et~al.}(2014)\citenamefont
  {Rakholia}, \citenamefont {McGuinness},\ and\ \citenamefont
  {Biedermann}}]{Rakholia2014}%
  \BibitemOpen
  \bibfield  {author} {\bibinfo {author} {\bibfnamefont {A.~V.}\ \bibnamefont
  {Rakholia}}, \bibinfo {author} {\bibfnamefont {H.~J.}\ \bibnamefont
  {McGuinness}}, \ and\ \bibinfo {author} {\bibfnamefont {G.~W.}\ \bibnamefont
  {Biedermann}},\ }\bibfield  {title} {\enquote {\bibinfo {title} {Dual-axis
  high-data-rate atom interferometer via cold ensemble exchange},}\ }\href
  {\doibase 10.1103/physrevapplied.2.054012} {\bibfield  {journal} {\bibinfo
  {journal} {Physical Review Applied}\ }\textbf {\bibinfo {volume} {2}}
  (\bibinfo {year} {2014}),\ 10.1103/physrevapplied.2.054012}\BibitemShut
  {NoStop}%
\bibitem [{\citenamefont {Dutta}\ \emph {et~al.}(2016)\citenamefont {Dutta},
  \citenamefont {Savoie}, \citenamefont {Fang}, \citenamefont {Venon},
  \citenamefont {{Garrido Alzar}}, \citenamefont {Geiger},\ and\ \citenamefont
  {Landragin}}]{Dutta2016}%
  \BibitemOpen
  \bibfield  {author} {\bibinfo {author} {\bibfnamefont {I.}~\bibnamefont
  {Dutta}}, \bibinfo {author} {\bibfnamefont {D.}~\bibnamefont {Savoie}},
  \bibinfo {author} {\bibfnamefont {B.}~\bibnamefont {Fang}}, \bibinfo {author}
  {\bibfnamefont {B.}~\bibnamefont {Venon}}, \bibinfo {author} {\bibfnamefont
  {C.~L.}\ \bibnamefont {{Garrido Alzar}}}, \bibinfo {author} {\bibfnamefont
  {R.}~\bibnamefont {Geiger}}, \ and\ \bibinfo {author} {\bibfnamefont
  {A.}~\bibnamefont {Landragin}},\ }\bibfield  {title} {\enquote {\bibinfo
  {title} {{Continuous Cold-Atom Inertial Sensor with 1 nrad/s Rotation
  Stability}},}\ }\href {\doibase 10.1103/PhysRevLett.116.183003} {\bibfield
  {journal} {\bibinfo  {journal} {Phys. Rev. Lett.}\ }\textbf {\bibinfo
  {volume} {116}},\ \bibinfo {pages} {183003} (\bibinfo {year}
  {2016})}\BibitemShut {NoStop}%
\bibitem [{Note1()}]{Note1}%
  \BibitemOpen
  \bibinfo {note} {Industry standards are 200 Hz for naval application and 2
  kHz for aviation}\BibitemShut {NoStop}%
\bibitem [{\citenamefont {Lautier}\ \emph {et~al.}(2014)\citenamefont
  {Lautier}, \citenamefont {Volodimer}, \citenamefont {Hardin}, \citenamefont
  {Merlet}, \citenamefont {Lours}, \citenamefont {{Pereira Dos Santos}},\ and\
  \citenamefont {Landragin}}]{Lautier2014}%
  \BibitemOpen
  \bibfield  {author} {\bibinfo {author} {\bibfnamefont {J.}~\bibnamefont
  {Lautier}}, \bibinfo {author} {\bibfnamefont {L.}~\bibnamefont {Volodimer}},
  \bibinfo {author} {\bibfnamefont {T.}~\bibnamefont {Hardin}}, \bibinfo
  {author} {\bibfnamefont {S.}~\bibnamefont {Merlet}}, \bibinfo {author}
  {\bibfnamefont {M.}~\bibnamefont {Lours}}, \bibinfo {author} {\bibfnamefont
  {F.}~\bibnamefont {{Pereira Dos Santos}}}, \ and\ \bibinfo {author}
  {\bibfnamefont {A.}~\bibnamefont {Landragin}},\ }\bibfield  {title} {\enquote
  {\bibinfo {title} {{Hybridizing matter-wave and classical accelerometers}},}\
  }\href {\doibase 10.1063/1.4897358} {\bibfield  {journal} {\bibinfo
  {journal} {Appl. Phys. Lett.}\ }\textbf {\bibinfo {volume} {105}},\ \bibinfo
  {pages} {144102} (\bibinfo {year} {2014})}\BibitemShut {NoStop}%
\bibitem [{\citenamefont {Ludlow}\ \emph {et~al.}(2015)\citenamefont {Ludlow},
  \citenamefont {Boyd}, \citenamefont {Ye}, \citenamefont {Peik},\ and\
  \citenamefont {Schmidt}}]{Ludlow2015}%
  \BibitemOpen
  \bibfield  {author} {\bibinfo {author} {\bibfnamefont {A.~D.}\ \bibnamefont
  {Ludlow}}, \bibinfo {author} {\bibfnamefont {M.~M.}\ \bibnamefont {Boyd}},
  \bibinfo {author} {\bibfnamefont {J.}~\bibnamefont {Ye}}, \bibinfo {author}
  {\bibfnamefont {E.}~\bibnamefont {Peik}}, \ and\ \bibinfo {author}
  {\bibfnamefont {P.~O.}\ \bibnamefont {Schmidt}},\ }\bibfield  {title}
  {\enquote {\bibinfo {title} {{Optical atomic clocks}},}\ }\href {\doibase
  10.1103/RevModPhys.87.637} {\bibfield  {journal} {\bibinfo  {journal} {Rev.
  Mod. Phys.}\ }\textbf {\bibinfo {volume} {87}},\ \bibinfo {pages} {637--701}
  (\bibinfo {year} {2015})}\BibitemShut {NoStop}%
\bibitem [{\citenamefont {Kalman}(1960)}]{Kalman1960}%
  \BibitemOpen
  \bibfield  {author} {\bibinfo {author} {\bibfnamefont {R.~E.}\ \bibnamefont
  {Kalman}},\ }\bibfield  {title} {\enquote {\bibinfo {title} {{A New Approach
  to Linear Filtering and Prediction Problems}},}\ }\href {\doibase
  10.1115/1.3662552} {\bibfield  {journal} {\bibinfo  {journal} {J. Basic
  Eng.}\ }\textbf {\bibinfo {volume} {82}},\ \bibinfo {pages} {35--45}
  (\bibinfo {year} {1960})}\BibitemShut {NoStop}%
\bibitem [{\citenamefont {Kalman}\ and\ \citenamefont
  {Bucy}(1961)}]{Kalman1961}%
  \BibitemOpen
  \bibfield  {author} {\bibinfo {author} {\bibfnamefont {R.~E.}\ \bibnamefont
  {Kalman}}\ and\ \bibinfo {author} {\bibfnamefont {R.~S.}\ \bibnamefont
  {Bucy}},\ }\bibfield  {title} {\enquote {\bibinfo {title} {{New Results in
  Linear Filtering and Prediction Theory}},}\ }\href {\doibase
  10.1115/1.3658902} {\bibfield  {journal} {\bibinfo  {journal} {J. Basic
  Eng.}\ }\textbf {\bibinfo {volume} {83}},\ \bibinfo {pages} {95--108}
  (\bibinfo {year} {1961})}\BibitemShut {NoStop}%
\bibitem [{\citenamefont {Bar-Shalom}\ \emph {et~al.}(2002)\citenamefont
  {Bar-Shalom}, \citenamefont {Li},\ and\ \citenamefont
  {Kirubarajan}}]{BarShalom2002}%
  \BibitemOpen
  \bibfield  {author} {\bibinfo {author} {\bibfnamefont {Y.}~\bibnamefont
  {Bar-Shalom}}, \bibinfo {author} {\bibfnamefont {X.-Rong}\ \bibnamefont
  {Li}}, \ and\ \bibinfo {author} {\bibfnamefont {T.}~\bibnamefont
  {Kirubarajan}},\ }\href {\doibase 10.1002/0471221279} {\emph {\bibinfo
  {title} {{Estimation with Applications to Tracking and Navigation: Theory,
  Algorithms and Software}}}}\ (\bibinfo  {publisher} {John Wiley {\&} Sons,
  Inc.},\ \bibinfo {address} {New York, NY, USA},\ \bibinfo {year}
  {2002})\BibitemShut {NoStop}%
\bibitem [{\citenamefont {{van Trees}}\ \emph {et~al.}(2013)\citenamefont {{van
  Trees}}, \citenamefont {Bell},\ and\ \citenamefont {Tian}}]{vanTrees2013}%
  \BibitemOpen
  \bibfield  {author} {\bibinfo {author} {\bibfnamefont {H.~L.}\ \bibnamefont
  {{van Trees}}}, \bibinfo {author} {\bibfnamefont {K.~L.}\ \bibnamefont
  {Bell}}, \ and\ \bibinfo {author} {\bibfnamefont {Z.}~\bibnamefont {Tian}},\
  }\href
  {https://www.amazon.com/Detection-Estimation-Modulation-Theory-Part/dp/0470542969}
  {\emph {\bibinfo {title} {{Detection, Estimation and Modulation Theory}}}},\
  \bibinfo {edition} {2nd}\ ed.\ (\bibinfo  {publisher} {John Wiley \& Sons,
  Inc.},\ \bibinfo {address} {Hoboken, NJ, USA},\ \bibinfo {year}
  {2013})\BibitemShut {NoStop}%
\bibitem [{\citenamefont {Brown}\ and\ \citenamefont
  {Hwang}(2012)}]{Brown2012}%
  \BibitemOpen
  \bibfield  {author} {\bibinfo {author} {\bibfnamefont {R.~G.}\ \bibnamefont
  {Brown}}\ and\ \bibinfo {author} {\bibfnamefont {P.~Y.~C.}\ \bibnamefont
  {Hwang}},\ }\href
  {https://www.amazon.com/Introduction-Signals-Applied-Filtering-Exercises/dp/0470609699?SubscriptionId=0JYN1NVW651KCA56C102&tag=techkie-20&linkCode=xm2&camp=2025&creative=165953&creativeASIN=0470609699}
  {\emph {\bibinfo {title} {{Introduction to Random Signals and Applied Kalman
  Filtering: with MATLAB Exercises}}}},\ \bibinfo {edition} {4th}\ ed.\
  (\bibinfo  {publisher} {John Wiley \& Sons, Inc.},\ \bibinfo {address}
  {Hoboken, NJ, USA},\ \bibinfo {year} {2012})\BibitemShut {NoStop}%
\bibitem [{\citenamefont {Merlet}\ \emph {et~al.}(2009)\citenamefont {Merlet},
  \citenamefont {{Le Gou{\"e}t}}, \citenamefont {Bodart}, \citenamefont
  {Clairon}, \citenamefont {Landragin}, \citenamefont {{Pereira Dos Santos}},\
  and\ \citenamefont {Rouchon}}]{Merlet2009}%
  \BibitemOpen
  \bibfield  {author} {\bibinfo {author} {\bibfnamefont {S.}~\bibnamefont
  {Merlet}}, \bibinfo {author} {\bibfnamefont {J.}~\bibnamefont {{Le
  Gou{\"e}t}}}, \bibinfo {author} {\bibfnamefont {Q.}~\bibnamefont {Bodart}},
  \bibinfo {author} {\bibfnamefont {A.}~\bibnamefont {Clairon}}, \bibinfo
  {author} {\bibfnamefont {A.}~\bibnamefont {Landragin}}, \bibinfo {author}
  {\bibfnamefont {F.}~\bibnamefont {{Pereira Dos Santos}}}, \ and\ \bibinfo
  {author} {\bibfnamefont {P.}~\bibnamefont {Rouchon}},\ }\bibfield  {title}
  {\enquote {\bibinfo {title} {{Operating an atom interferometer beyond its
  linear range}},}\ }\href {http://stacks.iop.org/0026-1394/46/i=1/a=011}
  {\bibfield  {journal} {\bibinfo  {journal} {Metrologia}\ }\textbf {\bibinfo
  {volume} {46}},\ \bibinfo {pages} {87} (\bibinfo {year} {2009})}\BibitemShut
  {NoStop}%
\bibitem [{\citenamefont {Barrett}\ \emph {et~al.}(2015)\citenamefont
  {Barrett}, \citenamefont {Antoni-Micollier}, \citenamefont {Chichet},
  \citenamefont {Battelier}, \citenamefont {Gominet}, \citenamefont {Bertoldi},
  \citenamefont {Bouyer},\ and\ \citenamefont {Landragin}}]{Barrett2015}%
  \BibitemOpen
  \bibfield  {author} {\bibinfo {author} {\bibfnamefont {B.}~\bibnamefont
  {Barrett}}, \bibinfo {author} {\bibfnamefont {L.}~\bibnamefont
  {Antoni-Micollier}}, \bibinfo {author} {\bibfnamefont {L.}~\bibnamefont
  {Chichet}}, \bibinfo {author} {\bibfnamefont {B.}~\bibnamefont {Battelier}},
  \bibinfo {author} {\bibfnamefont {P.-A.}\ \bibnamefont {Gominet}}, \bibinfo
  {author} {\bibfnamefont {A.}~\bibnamefont {Bertoldi}}, \bibinfo {author}
  {\bibfnamefont {P.}~\bibnamefont {Bouyer}}, \ and\ \bibinfo {author}
  {\bibfnamefont {A.}~\bibnamefont {Landragin}},\ }\bibfield  {title} {\enquote
  {\bibinfo {title} {{Correlative methods for dual-species quantum tests of the
  weak equivalence principle}},}\ }\href {\doibase
  10.1088/1367-2630/17/8/085010} {\bibfield  {journal} {\bibinfo  {journal}
  {New J. Phys.}\ }\textbf {\bibinfo {volume} {17}},\ \bibinfo {pages} {085010}
  (\bibinfo {year} {2015})}\BibitemShut {NoStop}%
\bibitem [{Note2()}]{Note2}%
  \BibitemOpen
  \bibinfo {note} {Nanometrics Titan force-balance accelerometer. Measurements
  were realized using the 0.5 g clip range.}\BibitemShut {Stop}%
\bibitem [{\citenamefont {Cheinet}\ \emph {et~al.}(2008)\citenamefont
  {Cheinet}, \citenamefont {Canuel}, \citenamefont {{Pereira Dos Santos}},
  \citenamefont {Gauguet}, \citenamefont {Yver-Leduc},\ and\ \citenamefont
  {Landragin}}]{Cheinet2008}%
  \BibitemOpen
  \bibfield  {author} {\bibinfo {author} {\bibfnamefont {P.}~\bibnamefont
  {Cheinet}}, \bibinfo {author} {\bibfnamefont {B.}~\bibnamefont {Canuel}},
  \bibinfo {author} {\bibfnamefont {F.}~\bibnamefont {{Pereira Dos Santos}}},
  \bibinfo {author} {\bibfnamefont {A.}~\bibnamefont {Gauguet}}, \bibinfo
  {author} {\bibfnamefont {F.}~\bibnamefont {Yver-Leduc}}, \ and\ \bibinfo
  {author} {\bibfnamefont {A.}~\bibnamefont {Landragin}},\ }\bibfield  {title}
  {\enquote {\bibinfo {title} {{Measurement of the Sensitivity Function in a
  Time-Domain Atomic Interferometer}},}\ }\href {\doibase
  10.1109/TIM.2007.915148} {\bibfield  {journal} {\bibinfo  {journal} {IEEE
  Trans. Instrum. Meas.}\ }\textbf {\bibinfo {volume} {57}},\ \bibinfo {pages}
  {1141--1148} (\bibinfo {year} {2008})}\BibitemShut {NoStop}%
\bibitem [{\citenamefont {Groves}(2013)}]{Groves2013}%
  \BibitemOpen
  \bibfield  {author} {\bibinfo {author} {\bibfnamefont {P.~D.}\ \bibnamefont
  {Groves}},\ }\href
  {https://www.amazon.com/Principles-Multisensor-Integrated-Navigation-Applications/dp/1608070050}
  {\emph {\bibinfo {title} {{Principles of GNSS, Inertial, and Multisensor
  Integrated Navigation Systems}}}},\ \bibinfo {edition} {2nd}\ ed.,\ GNSS
  Technology and Applications Series\ (\bibinfo  {publisher} {Artech House},\
  \bibinfo {address} {Norwood, MA, USA},\ \bibinfo {year} {2013})\BibitemShut
  {NoStop}%
\bibitem [{\citenamefont {Yonezawa}\ \emph {et~al.}(2012)\citenamefont
  {Yonezawa}, \citenamefont {Nakane}, \citenamefont {Wheatley}, \citenamefont
  {Iwasawa}, \citenamefont {Takeda}, \citenamefont {Arao}, \citenamefont
  {Ohki}, \citenamefont {Tsumura}, \citenamefont {Berry}, \citenamefont
  {Ralph}, \citenamefont {Wiseman}, \citenamefont {Huntington},\ and\
  \citenamefont {Furusawa}}]{Yonezawa2012}%
  \BibitemOpen
  \bibfield  {author} {\bibinfo {author} {\bibfnamefont {H.}~\bibnamefont
  {Yonezawa}}, \bibinfo {author} {\bibfnamefont {D.}~\bibnamefont {Nakane}},
  \bibinfo {author} {\bibfnamefont {T.~A.}\ \bibnamefont {Wheatley}}, \bibinfo
  {author} {\bibfnamefont {K.}~\bibnamefont {Iwasawa}}, \bibinfo {author}
  {\bibfnamefont {S.}~\bibnamefont {Takeda}}, \bibinfo {author} {\bibfnamefont
  {H.}~\bibnamefont {Arao}}, \bibinfo {author} {\bibfnamefont {K.}~\bibnamefont
  {Ohki}}, \bibinfo {author} {\bibfnamefont {K.}~\bibnamefont {Tsumura}},
  \bibinfo {author} {\bibfnamefont {D.~W.}\ \bibnamefont {Berry}}, \bibinfo
  {author} {\bibfnamefont {T.~C.}\ \bibnamefont {Ralph}}, \bibinfo {author}
  {\bibfnamefont {H.~M.}\ \bibnamefont {Wiseman}}, \bibinfo {author}
  {\bibfnamefont {E.~H.}\ \bibnamefont {Huntington}}, \ and\ \bibinfo {author}
  {\bibfnamefont {A.}~\bibnamefont {Furusawa}},\ }\bibfield  {title} {\enquote
  {\bibinfo {title} {{Quantum-Enhanced Optical-Phase Tracking}},}\ }\href
  {\doibase 10.1126/science.1225258} {\bibfield  {journal} {\bibinfo  {journal}
  {Science}\ }\textbf {\bibinfo {volume} {337}},\ \bibinfo {pages} {1514--1517}
  (\bibinfo {year} {2012})}\BibitemShut {NoStop}%
\bibitem [{\citenamefont {Jim\'enez-Mart\'{\i}nez}\ \emph
  {et~al.}(2018)\citenamefont {Jim\'enez-Mart\'{\i}nez}, \citenamefont
  {Ko\l{}ody\ifmmode~\acute{n}\else \'{n}\fi{}ski}, \citenamefont {Troullinou},
  \citenamefont {Lucivero}, \citenamefont {Kong},\ and\ \citenamefont
  {Mitchell}}]{Jimenez-Martinez2018}%
  \BibitemOpen
  \bibfield  {author} {\bibinfo {author} {\bibfnamefont {R.}~\bibnamefont
  {Jim\'enez-Mart\'{\i}nez}}, \bibinfo {author} {\bibfnamefont
  {J.}~\bibnamefont {Ko\l{}ody\ifmmode~\acute{n}\else \'{n}\fi{}ski}}, \bibinfo
  {author} {\bibfnamefont {C.}~\bibnamefont {Troullinou}}, \bibinfo {author}
  {\bibfnamefont {V.G.}\ \bibnamefont {Lucivero}}, \bibinfo {author}
  {\bibfnamefont {J.}~\bibnamefont {Kong}}, \ and\ \bibinfo {author}
  {\bibfnamefont {M.~W.}\ \bibnamefont {Mitchell}},\ }\bibfield  {title}
  {\enquote {\bibinfo {title} {{Signal Tracking Beyond the Time Resolution of
  an Atomic Sensor by Kalman Filtering}},}\ }\href {\doibase
  10.1103/PhysRevLett.120.040503} {\bibfield  {journal} {\bibinfo  {journal}
  {Phys. Rev. Lett.}\ }\textbf {\bibinfo {volume} {120}},\ \bibinfo {pages}
  {040503} (\bibinfo {year} {2018})}\BibitemShut {NoStop}%
\bibitem [{Note3()}]{Note3}%
  \BibitemOpen
  \bibinfo {note} {The measurement noise comprises the detection noise, the
  phase estimation noise and the true phase noise}\BibitemShut {NoStop}%
\bibitem [{Note4()}]{Note4}%
  \BibitemOpen
  \bibinfo {note} {For time-invariant driving and measurement, the steady-state
  covariance can be determined by solving the discrete-time algebraic Riccati
  equations. This is not possible here because the measurement and noise
  matrices $\protect \bm {H}$ and $\protect \bm {R}$ vary
  randomly.}\BibitemShut {Stop}%
\bibitem [{Note5()}]{Note5}%
  \BibitemOpen
  \bibinfo {note} {These data are acquired by a 16-bit acquisition system at a
  sampling rate of 50 kHz and averaged by packets of $5~\protect \mathrm {ms}$
  for the full 16 h duration of the dataset.}\BibitemShut {Stop}%
\bibitem [{Note6()}]{Note6}%
  \BibitemOpen
  \bibinfo {note} {The initial guess of each fit corresponds to the result of
  the previous one and the bias tracking is interpolated linearly between two
  consecutive stacks.}\BibitemShut {Stop}%
\bibitem [{\citenamefont {Camp}\ and\ \citenamefont
  {Vauterin}(2005)}]{Camp2005}%
  \BibitemOpen
  \bibfield  {author} {\bibinfo {author} {\bibfnamefont {M.~Van}\ \bibnamefont
  {Camp}}\ and\ \bibinfo {author} {\bibfnamefont {P.}~\bibnamefont
  {Vauterin}},\ }\bibfield  {title} {\enquote {\bibinfo {title} {Tsoft:
  graphical and interactive software for the analysis of time series and earth
  tides},}\ }\href {\doibase 10.1016/j.cageo.2004.11.015} {\bibfield  {journal}
  {\bibinfo  {journal} {Comput. Geosciences}\ }\textbf {\bibinfo {volume}
  {31}},\ \bibinfo {pages} {631--640} (\bibinfo {year} {2005})}\BibitemShut
  {NoStop}%
\bibitem [{\citenamefont {Canuel}\ \emph {et~al.}(2006)\citenamefont {Canuel},
  \citenamefont {Leduc}, \citenamefont {Holleville}, \citenamefont {Gauguet},
  \citenamefont {Fils}, \citenamefont {Virdis}, \citenamefont {Clairon},
  \citenamefont {Dimarcq}, \citenamefont {Bord\'e}, \citenamefont {Landragin},\
  and\ \citenamefont {Bouyer}}]{Canuel2006}%
  \BibitemOpen
  \bibfield  {author} {\bibinfo {author} {\bibfnamefont {B.}~\bibnamefont
  {Canuel}}, \bibinfo {author} {\bibfnamefont {F.}~\bibnamefont {Leduc}},
  \bibinfo {author} {\bibfnamefont {D.}~\bibnamefont {Holleville}}, \bibinfo
  {author} {\bibfnamefont {A.}~\bibnamefont {Gauguet}}, \bibinfo {author}
  {\bibfnamefont {J.}~\bibnamefont {Fils}}, \bibinfo {author} {\bibfnamefont
  {A.}~\bibnamefont {Virdis}}, \bibinfo {author} {\bibfnamefont
  {A.}~\bibnamefont {Clairon}}, \bibinfo {author} {\bibfnamefont
  {N.}~\bibnamefont {Dimarcq}}, \bibinfo {author} {\bibfnamefont {Ch.~J.}\
  \bibnamefont {Bord\'e}}, \bibinfo {author} {\bibfnamefont {A.}~\bibnamefont
  {Landragin}}, \ and\ \bibinfo {author} {\bibfnamefont {P.}~\bibnamefont
  {Bouyer}},\ }\bibfield  {title} {\enquote {\bibinfo {title} {{Six-Axis
  Inertial Sensor Using Cold-Atom Interferometry}},}\ }\href {\doibase
  10.1103/PhysRevLett.97.010402} {\bibfield  {journal} {\bibinfo  {journal}
  {Phys. Rev. Lett.}\ }\textbf {\bibinfo {volume} {97}},\ \bibinfo {pages}
  {010402} (\bibinfo {year} {2006})}\BibitemShut {NoStop}%
\bibitem [{\citenamefont {Wu}\ \emph {et~al.}(2017)\citenamefont {Wu},
  \citenamefont {Zi}, \citenamefont {Dudley}, \citenamefont {Bilotta},
  \citenamefont {Canoza},\ and\ \citenamefont {M\"{u}ller}}]{Wu2017}%
  \BibitemOpen
  \bibfield  {author} {\bibinfo {author} {\bibfnamefont {X.}~\bibnamefont
  {Wu}}, \bibinfo {author} {\bibfnamefont {F.}~\bibnamefont {Zi}}, \bibinfo
  {author} {\bibfnamefont {J.}~\bibnamefont {Dudley}}, \bibinfo {author}
  {\bibfnamefont {R.~J.}\ \bibnamefont {Bilotta}}, \bibinfo {author}
  {\bibfnamefont {P.}~\bibnamefont {Canoza}}, \ and\ \bibinfo {author}
  {\bibfnamefont {H.}~\bibnamefont {M\"{u}ller}},\ }\bibfield  {title}
  {\enquote {\bibinfo {title} {{Multiaxis atom interferometry with a
  single-diode laser and a pyramidal magneto-optical trap}},}\ }\href {\doibase
  10.1364/OPTICA.4.001545} {\bibfield  {journal} {\bibinfo  {journal} {Optica}\
  }\textbf {\bibinfo {volume} {4}},\ \bibinfo {pages} {1545--1551} (\bibinfo
  {year} {2017})}\BibitemShut {NoStop}%
\bibitem [{\citenamefont {Gauguet}\ \emph {et~al.}(2009)\citenamefont
  {Gauguet}, \citenamefont {Canuel}, \citenamefont {L\'ev\`eque}, \citenamefont
  {Chaibi},\ and\ \citenamefont {Landragin}}]{Gauguet2009}%
  \BibitemOpen
  \bibfield  {author} {\bibinfo {author} {\bibfnamefont {A.}~\bibnamefont
  {Gauguet}}, \bibinfo {author} {\bibfnamefont {B.}~\bibnamefont {Canuel}},
  \bibinfo {author} {\bibfnamefont {T.}~\bibnamefont {L\'ev\`eque}}, \bibinfo
  {author} {\bibfnamefont {W.}~\bibnamefont {Chaibi}}, \ and\ \bibinfo {author}
  {\bibfnamefont {A.}~\bibnamefont {Landragin}},\ }\bibfield  {title} {\enquote
  {\bibinfo {title} {{Characterization and limits of a cold-atom Sagnac
  interferometer}},}\ }\href {\doibase 10.1103/PhysRevA.80.063604} {\bibfield
  {journal} {\bibinfo  {journal} {Phys. Rev. A}\ }\textbf {\bibinfo {volume}
  {80}},\ \bibinfo {pages} {063604} (\bibinfo {year} {2009})}\BibitemShut
  {NoStop}%
\bibitem [{\citenamefont {Stockton}\ \emph {et~al.}(2007)\citenamefont
  {Stockton}, \citenamefont {Wu},\ and\ \citenamefont
  {Kasevich}}]{Stockton2007}%
  \BibitemOpen
  \bibfield  {author} {\bibinfo {author} {\bibfnamefont {J.~K.}\ \bibnamefont
  {Stockton}}, \bibinfo {author} {\bibfnamefont {X.}~\bibnamefont {Wu}}, \ and\
  \bibinfo {author} {\bibfnamefont {M.~A.}\ \bibnamefont {Kasevich}},\
  }\bibfield  {title} {\enquote {\bibinfo {title} {{Bayesian estimation of
  differential interferometer phase}},}\ }\href {\doibase
  10.1103/PhysRevA.76.033613} {\bibfield  {journal} {\bibinfo  {journal} {Phys.
  Rev. A}\ }\textbf {\bibinfo {volume} {76}},\ \bibinfo {pages} {033613}
  (\bibinfo {year} {2007})}\BibitemShut {NoStop}%
\bibitem [{\citenamefont {Bonnin}\ \emph {et~al.}(2013)\citenamefont {Bonnin},
  \citenamefont {Zahzam}, \citenamefont {Bidel},\ and\ \citenamefont
  {Bresson}}]{Bonnin2013}%
  \BibitemOpen
  \bibfield  {author} {\bibinfo {author} {\bibfnamefont {A.}~\bibnamefont
  {Bonnin}}, \bibinfo {author} {\bibfnamefont {N.}~\bibnamefont {Zahzam}},
  \bibinfo {author} {\bibfnamefont {Y.}~\bibnamefont {Bidel}}, \ and\ \bibinfo
  {author} {\bibfnamefont {A.}~\bibnamefont {Bresson}},\ }\bibfield  {title}
  {\enquote {\bibinfo {title} {{Simultaneous dual-species matter-wave
  accelerometer}},}\ }\href {\doibase 10.1103/PhysRevA.88.043615} {\bibfield
  {journal} {\bibinfo  {journal} {Phys. Rev. A}\ }\textbf {\bibinfo {volume}
  {88}},\ \bibinfo {pages} {043615} (\bibinfo {year} {2013})}\BibitemShut
  {NoStop}%
\end{thebibliography}
\end{document}